\documentclass[useAMS,usenatbib,referee]{mn2e}
\def \degmark{^\circ}

\def \arcmin {\hbox{$^\prime$}}

\def \blfootnote{\xdef\@thefnmark[]{}\footnotetext[]} 

\def \sdsssource {\hbox{SDSS~J123932.75+044305.3}}
\def \gbsource {\hbox{GB6~J1239+0443}}
\def \egretsource {\hbox{3EG~J1236+0457}}
\def \aglsource {\hbox{AGL~J1238+0406}}
\def \fermisource {\hbox{2FGL J1239.5+0443}}
\def \bzsource {\hbox{BZQ~1239+0443}}

\def \b454 {\hbox{3C~454.3}}

\newcommand\ion[2]{#1$\;${\sc{#2}}}%

\usepackage{epsfig}

\title[Flaring and low activity periods of \gbsource]
{The characterization of the distant blazar \gbsource\ from flaring and low activity periods}

\author[L. Pacciani et al.]{L. Pacciani$^1$,
I. Donnarumma$^1$,
K. D. Denney$^{2}$, R. J. Assef$^{3}$, Y. Ikejiri$^{4}$,
\newauthor
M. Yamanaka$^{4}$,
M. Uemura$^{5}$,
A. Domingo$^{6}$,
P. Giommi$^{7}$,
A. Tarchi$^{8}$,
\newauthor
F. Verrecchia$^{7}$,
F. Longo$^{9}$,
S. Rain{\'o}$^{10}$,
M. Giusti$^1$,
S. Vercellone$^{11}$,
A. W. Chen$^{12}$,
\newauthor
E. Striani$^1$,
V. Vittorini$^{1,13}$,
M. Tavani$^{1,13}$,
A. Bulgarelli$^{14}$,
A. Giuliani$^{12}$,
\newauthor
G. Pucella$^{15}$,
A. Argan$^1$,
G. Barbiellini$^{9}$,
P. Caraveo$^{12}$,
P. W. Cattaneo$^{16}$,
\newauthor
S. Colafrancesco$^{17,18}$,
E. Costa$^1$,
G. De Paris$^1$,
E. Del Monte$^1$,
G. Di Cocco$^{19}$,
\newauthor
Y. Evangelista$^1$,
A. Ferrari$^{20}$, M. Feroci$^1$,
M. Fiorini$^{12}$,
F. Fuschino$^{19}$,
M. Galli$^{21}$,
\newauthor
F. Gianotti$^{19}$,
C. Labanti$^{19}$, I. Lapshov$^1$, F. Lazzarotto$^1$,
P. Lipari$^{22}$,
M. Marisaldi$^{19}$,
\newauthor
S. Mereghetti$^{12}$, E. Morelli$^{19}$, E. Moretti$^{9}$,
A. Morselli$^{23}$,
A. Pellizzoni$^{8}$,
\newauthor
F. Perotti$^{12}$,
G. Piano$^{1,13,23}$, P. Picozza$^{13,23}$,
M. Pilia$^{24,8}$,
M. Prest$^{24}$,
\newauthor
M. Rapisarda$^{15}$,
A. Rappoldi$^{16}$,
A. Rubini$^1$,
S. Sabatini$^1$,
P. Soffitta$^1$,
\newauthor
M. Trifoglio$^{19}$, A. Trois$^{8}$, E. Vallazza$^{9}$,
D. Zanello$^{22}$,
C. Pittori$^{7}$,
\newauthor
P. Santolamazza$^{7}$, F. Lucarelli$^{7}$,
L. Salotti$^{14}$
and G. Valentini$^{14}$\\
$^{1}$INAF/IAPS, Via Fosso del Cavaliere, 100 I-00133 Roma, Italy \\
$^{2}$Marie Curie Fellow, Dark Cosmology Center, Niels Bohr Institute, University of Copenhagen, Copenhagen, Denmark\\
$^{3}$NASA Postdoctoral Program fellow at the Jet Propulsion Laboratory, California Institute of Technology\\
MS 169-236, 4800 Oak Grove Dr., Pasadena, 91109\\
$^{4}$Department of Physical Science, Hiroshima University, 1-3-1 Kagamiyama, Higashi-Hiroshima 739-8526, Japan\\
$^{5}$Hiroshima Astrophysical Science Center, Hiroshima University, 1-3-1 Kagamiyama, Higashi-Hiroshima 739-8526, Japan\\
$^{6}$Centro de Astrobiolog{\'{\i}}a (INTA-CSIC). PO Box 78, 28691 Villanueva de la Ca\~{n}ada, Madrid, Spain\\
$^{7}$ASI Science Data Center, Via Galileo Galilei, I-00044 Frascati (Roma), Italy\\
$^{8}$INAF-OAC, localita' Poggio dei Pini, strada 54, I-09012 Capoterra (CA), Italy\\
$^{9}$Dip. Fisica, Univ Trieste and INFN Trieste, Via A. Valerio, 2, I-34127 Trieste, Italy\\
$^{10}$Istituto Nazionale di Fisica Nucleare, Sezione di Bari, Via Orabona 4, I-70125, Bari, Italy\\
$^{11}$INAF-IASF Palermo, Via Ugo La Malfa 153, I-90146 Palermo, Italy\\
$^{12}$INAF/IASF-Milano, Via E. Bassini, 15 I-20133 Milano, Italy \\
$^{13}$Dip. di Fisica, Univ. Tor Vergata, Via della Ricerca Scientifica, 1 I-00133 Roma, Italy \\
$^{14}$Agenzia Spaziale Italiana, Viale Liegi, 26 I-00198 Roma, Italy \\
$^{15}$ENEA Frascati,  Via Enrico Fermi, 13 I-00044 Frascati (Roma), Italy \\
$^{16}$INFN-Pavia, Via Agostino Bassi, 6, I-27100 Pavia, Italy \\
$^{17}$INAF-OAR, Via di Frascati, 33 I-00040, Monteporzio Catone (Roma), Italy\\
$^{18}$School of Physics, University of the Witwatersrand, Johannesburg Wits 2050, South Africa.
$^{19}$INAF/IASF-Bologna, Via Gobetti 101, I-40129 Bologna, Italy \\
$^{20}$Dip. Fisica, Universit\'a di Torino, Via Giuria, 1, I-10125, Torino, Italy \\
$^{21}$ENEA-Bologna, Via Martiri Montesole, 4 I-40129 Bologna, Italy\\
$^{22}$INFN-Roma La Sapienza, P.le A. Moro, 2 I-00185 Roma, Italy \\
$^{23}$INFN Roma Tor Vergata, Via della Ricerca Scientifica, 1 I-00133 Roma, Italy\\
$^{24}$Dip. di Fisica, Univ. Dell'Insubria, Via Valleggio 11, I-22100 Como, Italy
}
\date{Submitted 2011 November 10}
\pagerange{\pageref{firstpage}--\pageref{lastpage}} \pubyear{2012}
\def\LaTeX{L\kern-.36em\raise.3ex\hbox{a}\kern-.15em
    T\kern-.1667em\lower.7ex\hbox{E}\kern-.125emX}

\begin{document}
\label{firstpage}
\maketitle
\newpage
\begin{abstract}
In 2008 {\it AGILE} and {\it Fermi} detected gamma-ray flaring activity from the unidentified EGRET source \egretsource,
recently  associated with a flat spectrum radio quasar (\gbsource) at z=1.762.
The optical counterpart of the gamma-ray source underwent
%GG%
a flux enhancement of a factor 15-30 in 6 years, and of 
$\sim$10 in six months.
We interpret this flare-up in terms of a transition from an
accretion-disk dominated emission to a synchrotron-jet dominated one.\\
%GG%
We analysed a Sloan Digital Sky Survey (SDSS) archival optical spectrum 
taken during a period of low radio and optical activity of the source. We 
estimated the mass of the central black hole using the  width of 
the C\,{\sc iv} emission line. In our work, we have also investigated SDSS archival    
optical photometric data and UV GALEX observations to estimate the 
thermal-disk emission contribution of \gbsource.\\
%GG%
Our analysis of the gamma-ray data taken during the flaring episodes 
indicates a flat gamma-ray spectrum, with an extension of up to 15
GeV, with no statistically-relevant sign of absorption from the broad
line region, suggesting that the blazar-zone is located beyond
the broad line region. This result is confirmed by the modeling of the 
broad-band spectral energy distribution (well constrained by the 
available multiwavelength data) of the flaring activity periods and by the 
accretion disk luminosity and black hole mass estimated by us using archival 
data.
%GG%
\end{abstract}

%GG%\keywords{ galaxies: active - galaxies: quasars: general - galaxies: individual: \gbsource\ - radiation mechanism: non thermal}
\section{Introduction}
Blazars are a sub-class of active galactic nuclei, emitting from radio to TeV energies.
They are subdivided in two main categories: Flat Spectrum Radio Quasars (FSRQ) and BL Lacertae (BL Lac) objects.
FSRQs are characterized by a flat radio spectrum in the GHz range (with spectral index $\alpha\le0.5$, where the flux density is $S_\nu\propto\nu^{-\alpha}$), and strong and broad emission lines (with equivalent width $>$5 $\AA$).
BL Lac objects, on the other hand, have no or weak emission lines with equivalent width $<$5 $\AA$.\\
Blazars continuum emission originates from a relativistic jet aligned with the line of sight.
Their Spectral Energy Distribution (SED) shows a double humped shape \citep{padovani1995}, with a low energy peak lying between IR and X-rays, and an high energy peak in the MeV-TeV band.\\
The low energy region of blazar spectra is associated with the synchrotron emission coming from the jet relativistic electrons. The high energy region can be modeled through
inverse Compton emission (leptonic models), with seed photons coming from an external region (e.g. the accretion disk, the dusty torus),
eventually reprocessed by the broad line region, or the hot corona, or from the synchrotron
process itself (synchrotron self-Compton, or SSC). A detailed description of leptonic models can be found in \cite{maraschi1992,marscher1992,sikora1994}.

The high energy region can also be modeled with hadronic scenarios \citep{mucke2001,mucke2003,bottcher2007}, where the very high energy protons of the jet are radiatively important. 
The accelerated protons produce gamma ray emission through proton synchrotron emission, the decay of neutral pions, and synchrotron emission produced by secondaries.\\
%
%GG%The high energy region can also be modeled within the hadronic scenario , where the very high energy protons of the jet result to be the emission key elements:
%GG%the accelerated protons produce proton-synchrotron emission, synchrotron through pair production, and gamma-ray emission through the decay of neutral pion produced in the proton initiated cascades.
%GG%The cascades may originate from high energy protons (above pion-production threshold) and synchrotron photons ({\it p-synchrotron cascade}).}\\
%GG%
The location of the so-called ``blazar-zone'', i.e., the spatial
location of the blazar SED-peaks and gamma-ray emitting region, in
blazars is still a matter of debate. 
\cite{sikora2008} proposed that the blazar-zone is located at 3--9 pc
from the central engine for the outburst of 3C~454.3 (a bright FSRQ) occurred in
2005. For the same flare, \cite{ghisellini2007} indicated, instead, a
dissipation region at 0.5--0.8 pc from the central black hole (BH).
From the combined study of time-dependent polarimetric radio images at 43 and 86 GHz, the optical polarimetry, and radio, optical, X-, gamma-ray light curves,
\cite{jorstad2010} proposed that the low and high energy emission is
located near the 43 GHz core, at a distance of the order of tens of
parsecs from the BH for 3C~454.3. A similar investigation,  performed 
for the BL Lac objects OJ287 and AO~0235+164 \citep{agudo2011a,agudo2011b},
led to similar results.\\
\cite{tavecchio2009} established that gamma-rays emitted inside the broad line region (BLR) are absorbed at E$> 20$ $\mathrm{GeV}/(1+z)$
due to the $\gamma \gamma$ interaction  with the BLR photons (internal absorption). \cite{poutanen2010} refined this result, and claimed internal absorption features
at E$>5$ $\mathrm{GeV}$/(1+z) and at E$> 20$ $\mathrm{GeV}/(1+z)$ in the gamma-ray spectrum of 3C~279, 3C~454.3, PKS~1510-08, and a few other FSRQ.\\
%
%GG%If the gamma-rays are
%GG%emitted inside the broad line region (BLR), absorption at E$> 20\mathrm{GeV}/(1+z)$ {\bf is assumed to be} \citep{tavecchio2009} or at E$>5$
%GG%$\mathrm{GeV}$/(1+z) \citep{poutanen2010}, due to the $\gamma \gamma$
%GG%interaction of the gamma-rays coming from the jet with the BLR photons
%GG%(i.e., internal absorption).
%GG%Signatures of internal absorption have been
%GG%reported only for  3C~279, 3C~454.3, PKS~1510-08,
%GG%and a few other blazars \citep{poutanen2010}.
Within the leptonic scenario, \cite{ghisellini2009} show that the
contribution of external photon fields, including contributions from the
BLR and dusty torus to the inverse Compton emission can be parametrized
as a function of the accretion disk luminosity and the dissipation
distance of the emitting blob from the BH \citep{ghisellini2009}.
\cite{ghisellini2010} modeled the SED of the gamma-ray loudest blazars 
as being emitted at 0.01--0.5 pc from the BH.\\
%GG%
%GG%The
%GG%SED of the gamma-ray loudest blazars have thus
%GG%been modelled as being emitted at 0.01--0.5 pc from the BH
%GG%\citep{ghisellini2010}.
%

Using multiwavelength observations of the blazar \gbsource, we will
apply the models of \cite{ghisellini2009} to further investigate the
location of the blazar-zone.\\
\gbsource\ was an unidentified gamma-ray
source of the Virgo region (\egretsource), detected with low
significance \citep{hartman1999,casandjian2008} by the EGRET gamma-ray telescope (operating in the 30 MeV -- 30 GeV energy range, see \citealt{esposito1999}).
%GG%, based on the full mission data-set. 
In recent years the gamma-ray source has shown two episodes
of remarkable high energy activity: at the beginning and at the end of
2008, when it was detected by the {\it AGILE} \citep{pacciani2009} and the {\it Fermi}--LAT \citep{tramatel} gamma-ray telescopes, respectively;
then named \aglsource\ \citep{pittori2009}, and \fermisource\ \citep{fermicat2}.
The accurate source location determined by {\it Fermi}--LAT allowed for the association of
the unidentified gamma-ray source with \bzsource\, a flat
spectrum radio quasar (FSRQ) included into the second edition of the
Roma-BZCAT Multi-Frequency Catalog of Blazars \citep{massaro2010}.  The
optical counterpart of this source is \sdsssource, located at z=1.762,
and the radio counterpart is named \gbsource.  In the following sections, we
refer to this object as \gbsource, its radio source name.\\
Here we present the results of an analysis of multifrequency data
simultaneous to the {\it AGILE} campaign on the Virgo field, and to the
follow-up carried out after the {\it Fermi}--LAT detection and localization.
By analysing the archival data, we estimate some fundamental physical
properties of the source such as the accretion disk luminosity and the
supermassive black hole (SMBH) mass. 
In this way we can obtain a consistent picture of the source emission in
the framework of leptonic models of blazars for periods of both low and
high emission activity.
In Sections 2 and 3 we will describe the multi-wavelength campaigns related to this source.
In Section 4 we will report on the archival data.
In section 5 we will present our results consisting in the determination of the BH mass, the gamma-ray light curve and spectrum, and the SED modeling.
\section{Gamma-ray observations and related multifrequency campaigns}
The {\it AGILE}--GRID gamma-ray telescope (operating in the 30 MeV -- 50 GeV energy range, see \citealt{agile}) performed two observing campaigns of the Virgo
field. The first campaign included 3 observations from 2007 December
16 to 2008 January 8. There were 3 simultaneous observations
(revolutions 633, 635, 637) with the wide-field instruments  aboard the {\it
 INTEGRAL} mission (operating in optical, hard X-, and soft gamma-rays, see \citealt{winkler2003}).  During this campaign {\it AGILE}
detected high gamma-ray activity from \gbsource\ \citep{pacciani2009}.
In the following sections, we will refer to this time period as {\em
  period A}, and to the related multiwavelength campaign as {\em
  campaign A}.
\\
{\it AGILE} also observed the Virgo field from 2009 June 4 to 2009
June 15, but there were no simultaneous observations with
wide-field instruments operating at other wavelengths.
\gbsource\ was undetected during this observation.\\
The {\it Fermi}--LAT gamma-ray telescope (20 MeV -- 300 GeV, see \citealt{atwood2009}), operating in all-sky survey mode since
2008 August 4, detected high gamma-ray activity at the end of December
2008 \citep{tramatel}, and triggered optical-UV/X-ray observations with
the Swift satellite \citep{Gehrels2004}, starting from 2009 January 2 for a total exposure of 4.7~ks.  Optical V
band photometry and polarization measurements were also made on-ground with the KANATA telescope
on 2009 January 2 and on 2009 January 3 \citep{trispecatel}.
In the following sections, we will refer to the observations collected during this
period as {\em period B}, and to the related multiwavelength campaign as
{\em campaign B}.
A summary of the observations is provided in Table \ref{tab:obs}.
\begin{table*}
\caption{Observing campaigns of the Virgo field during high gamma-ray activity periods from \gbsource.}
\label{tab:obs}
%\centering
\begin{tabular}{c | c | c }
\hline\hline
campaign        & observatory &  observing period                         \\ \hline
                &             & from 2007 Dec. 16 17:14 to 2007 Dec 23 02:18 UT\\ \cline{3-3}
                &  {\it AGILE}      & from 2007 Dec. 24 07:12 to 2007 Dec. 30 23:03 UT\\ \cline{3-3}
A               &             & from 2008 Jan. 04 13:35 to 2008 Jan. 08 11:06 UT\\ \cline{2-3}
                &             & from 2007 Dec. 19 18:08 to 2007 Dec. 22 06:44 UT (revolution 633)\\ \cline{3-3}
                &  {\it INTEGRAL}   & from 2007 Dec. 25 17:39 to 2007 Dec. 28 06:27 UT (revolution 635)\\ \cline{3-3}
                &             & from 2007 Dec. 31 17:13 to 2008 Jan. 03 04:00 UT (revolution 637)\\ \hline
                &  {\it Fermi}      & Dec. 2008 -- Jan. 2009                        \\ \cline{2-3}
B               &  Swift      & from 2009 Jan. 02 17:47 to  2009 Jan. 04 20:21 UT$^*$ (total exposure 4.7ks) \\ \cline{2-3}
                &  KANATA     & 2009 Jan. 02 18:14 UT for 1.5ks               \\ \cline{3-3} 
                &             & 2009 Jan. 03 19:41 UT for 1.5ks               \\ \hline
\end{tabular}
\\$^*$ Two pointings on 2009 Jan. 02 (exposure 2562 s and 380 s),  
and 5 pointings on 2009 Jan. 04 (exposure 597 s, 537s , 537 s, 577 s, 537 s)
\end{table*}
\section{Data analysis}
\subsection{{\it AGILE}-GRID data}

Level-1 {\it AGILE}-GRID data for campaign A were analysed using the
{\it AGILE} Standard Analysis Pipeline (BUILD20) and the {\it AGILE}
Scientific Analysis Package, based on the likelihood method
\citep{mattox1996}.  Albedo photons were rejected by applying a cut at
85$^\circ$ centered on the Earth. We selected well-reconstructed
gamma-rays by applying the FM3.119 filter, calibrated in the 100 MeV --
3 GeV energy band \citep{cattaneo2011}.  All the events collected during
the passage in the South-Atlantic Anomaly were rejected.  Counts,
exposure and Galactic background gamma-ray maps were created with
a bin-size of 0$^\circ$.1$\times$0$^\circ$.1 for E $>$ 100 MeV.  We
detected a source (\aglsource\ in the {\it AGILE} catalog, see
\citealt{pittori2009}, and \citealt{verrecchia2011}) with a significance
of $\sim$6 \citep[as measured by the $\sqrt{TS}$ parameter, see
][]{mattox1996}, located at $\alpha_{2000}$=190.25,
$\delta_{2000}$=4.40, with an error radius of (33 + 6) arcmin
(statistical error at 95\% confidence level -C.L.-, and systematic error, respectively), by
integrating the GRID data for 4 days between 2008 January 4 13:35 and
2008 January 8 11:16 (within the campaign A).  The source was
positionally consistent with \gbsource.
%GG%with an angular separation of 29 arcmin between the two localizations.
A multisource maximum likelihood analysis was performed to extract the
source flux and position taking into account nearby sources 3C 273, 3C 279, and 4C 04.42
(for which we obtained  $\sqrt{TS}$ $>$ 1 from a preliminary analysis
of the observing campaign).
For \aglsource\ we obtained a flux of (62$\pm$9)$\times 10^{-8}$
ph cm$^{-2}$ s$^{-1}$ (E $>$ 100 MeV) and a photon index 1.92$\pm$0.14.
The integration of the first week of observations with {\it AGILE} gave
no detection at the position of \gbsource, resulting in an upper
limit of 21$\times$10$^{-8}$ ph cm$^{-2}$ s$^{-1}$.
%GG%The {\it AGILE} light curve is reported in Fig. \ref{fig:lc} together with the {\it Fermi}--LAT data.
\subsection{{\it Fermi}--LAT data}
We analysed the {\it Fermi} Large Area Telescope ({\it Fermi}--LAT) data
for campaign B with the standard {\it Fermi} Science tools v9r23p1,
following the prescriptions in the online documentation
\footnote{http://fermi.gsfc.nasa.gov/ssc/data/analysis/documentation/}.
We used the Pass 7 response functions (P7\_V6). In particular, we
selected events of {\em event class} 2, suitable for point-like sources,
and we filtered out photons from the Earth's limb with a cut at
100$^\circ$ in the zenith angle.  We performed the unbinned likelihood
analysis inside a region of radius 15$^\circ$ around \gbsource\ to
derive the source flux.  We took into account the diffuse backgrounds,
which were modeled using gal\_2yearp7v6\_v0 for the Galactic diffuse
emission and iso\_p7v6source for the extragalactic isotropic emission
models\footnote{both available on the URL http://fermi.gsfc.nasa.gov/ssc/data/access/lat/BackgroundModels.html},
and all the 51 point-like gamma-ray sources in the second {\it
  Fermi}--LAT catalog \citep{fermicat2} within a slightly larger radius
of 20$^\circ$ from \gbsource\ (we considered a larger radius due to the
PSF width). For each source we used the model specified in the second
{\it Fermi}--LAT catalog.  For the sources within 10$^\circ$ from
\gbsource, we kept free all the spectral parameters in the fit. For the
sources within an annulus of internal radius 10$^\circ$ and external
radius 15$^\circ$ we kept free only the parameters related to the flux
normalization, and all the other parameters were fixed to the values
reported in the second {\it Fermi}--LAT catalog. For the sources outside
15$^\circ$ from \gbsource, we fixed all the spectral parameters to the
values reported in the catalog. This is a standard procedure for the
analysis of {\it Fermi}--LAT data, implemented with the python routine
{\em make2FGLxml.py}\footnote{available on the URL http://fermi.gsfc.nasa.gov/ssc/data/analysis/user/}
(contributed software by T. Johnson)\\
We proceeded with the analysis for energies only above 300 MeV, in order
to process data with a smaller point spread function and reduce
background gamma-rays from 3C 273, a bright and soft gamma-ray source
($\Gamma\sim 2.6$) located at $\sim$4$^{\circ}$ from \gbsource, since
the 68\% (95\%) containment radius for {\it Fermi}--LAT at normal
incidence is 4$^\circ$.5 (10$^\circ$) at 100 MeV
\footnote{http://www.slac.stanford.edu/exp/glast/groups/canda/archive/lat\_Performance.htm}.
Our study showed that this choice had a negligible effect on the
signal significance of \gbsource.\\
We used the {\em gtfindsrc} tool to locate the gamma-ray source. By
integrating data for 1 month centered around the peak flux (2008
December 29 16:00 UT)
%GG%from 2008 December 21~9:00 to 2009 January 4~9:00
in the band 300 MeV - 20 GeV, and by using photons converted both in the
front- and back-section of the {\it Fermi}--LAT, we obtained a detection
of a source (\fermisource\ in the second {\it Fermi}--LAT catalog, see
\citealt{fermicat2}) with $\sqrt{TS}$=20 located at
$\alpha_{2000}$=189.897, $\delta_{2000}$=4.718 and an error radius of 9
arcmin. The source was positionally consistent with \gbsource.
We obtained a gamma-ray photon index of 2.15 $\pm$ 0.11.\\
From the analysis in the 300 MeV -- 20 GeV range, we also obtained a
detection with $ \sqrt{TS}\sim$18, a flux of (23$\pm$3)$\times 10^{-8}$
ph cm$^{-2}$ s$^{-1}$, and a photon index of 2.21 $\pm$ 0.15, when
keeping the integration time within only 4 days centered at 2008
December 29 16:00 UT (the peak flux).  During this integration period,
the source was detected up to the energy interval 10--20 GeV, for which
we obtained a detection with $\sqrt{TS}=5.8$.
%gg%In particular, the unbinned likelihood analysis gives $\sim$2 gamma-rays from the \gbsource, and $\sim$6 gamma-rays from extragalactic+galactic background, and almost nothing from the other sources.\\ 
To compute the upper limits needed to build the source light curve and
spectra, we used the {\em UpperLimits} python function provided with the
{\it Fermi} Science tools.

\subsection{{\it INTEGRAL}/IBIS data}
The {\it INTEGRAL}/IBIS (operating in the 17 -- 400 keV energy range, see \citealt{ubertini2003}) data for campaign A were
processed using the OSA software version 8.0. We searched for the source
starting from the images accumulated in the 20-40 keV band for
revolutions 633, 635, 637 (simultaneous with {\it AGILE} observation of
the Virgo field, campaign A).  IBIS did not detect the source. We
derived a 3 sigma upper limit of 1.7~mCrab for each revolution
(200 ks exposure).

\subsection{Swift-XRT data}
The \emph{Swift}-XRT (X-ray Telescope, operating in the 0.2 -- 10 keV range, see \citealt{burrows2005}) data for campaign B were
processed using the most recent available calibration files. We made use
of \emph{Swift} Software version 3.5, FTOOLS version 6.8, and XSPEC
version 12.5.  The observations were obtained in photon counting mode,
with a total integration of 4.7~ks.  The mean source count rate was
(2.58$\pm$0.23)$\times$10$^{-2}$~cps.  We extracted the spectrum using a
photon binning ratio that ensured more than 20 photon counts per energy
bin.
%GG%This typically resulted in more than 55 energy bins in the spectrum and dropped down to a minumum of 20 energy bins per observation
%GG%for the smallest snapshots.
%GG%, which is more than sufficient to contrain the spectral parameters.
%GG%Once the spectral fits were obtained,
%GG%the flux was calculated in the 0.3-10 keV range for the light curve.\\
%
We fitted the X-ray data with an absorbed power law, fixing the absorption to the galactic value of 1.85$\times$10$^{20}$ cm$^{-2}$ \citep{dickey1990}.
We obtained a photon index of 1.42$\pm$0.25 (90\% C.L.). The estimated flux in the range 2-10~keV was (8.8$\pm$2.7)$\times$10$^{-13}$ erg cm$^{-2}$ s$^{-1}$ (68\% C.L.).
\subsection{Swift-UVOT data}
\textit{Swift}-UVOT (Ultra-Violet/Optical Telescope, see \citealt{roming2005}) data from each observation sequence of period B were
processed by the standard UVOT tool  \texttt{uvotsource} using the
same version of the Swift software as for the XRT analysis. An
extraction region of radius 5 arcsec centered on the source and a
background region of radius 13 arcsec located at
$\alpha_{2000}=$12$^{h}$39$^{m}$29$\fs$66,
$\delta_{2000}$=+04$\degmark$$42$'$34\farcs$ 2 (at least 27 arcsec far away from any object in the NED database)
were used.
Magnitudes are expressed in the
UVOT photometric system \citep{Poole08}. We obtained $m_U$=16.27$\pm$0.03 for \gbsource\ (extinction corrected
using the mean Galactic interstellar extinction curve from \citealt{Fitzpatrick1999}).
%GG%We checked the exposure and flat-field correction with two nearby reference stars SDSS J123939.08+044330.5 ($m_U$=18.99$\pm$0.02) and SDSS J123930.08+043953.1 (with $m_U$=15.17).
%
\subsection{Kanata optical data}
The optical photometry was performed using TRISPEC (a simultaneous optical and near-infrared imager, spectrograph, and polarimeter, see \citealt{watanabe2005})
attached to the Kanata 1.5-m telescope at Higashi-Hiroshima Observatory
on 2009 January 2 at 18:14 and 2009 January 3 at 19:41 UT (period B).
The observations were performed in \emph{polarimetry-mode} with a narrow
aperture mask of 1.5' width.  The total exposure was 1476~s per night.
The observations were pipeline-reduced, including bias removal and
flat-field corrections. We derived the V-band magnitude from
differential photometry with a nearby reference star at
$\alpha_{2000}=$12$^{h}$39$^{m}$30$\fs$11,
%GG%$\delta_{2000}=$+04$\arcdeg$39$'$52\farcs 6 
$\delta_{2000}$=+04$\degmark$$39$'$52\farcs$ 6 
of which the $V$ magnitude of
$14.095$ was deduced from the $g'$ and $r'$ data in the 6$^{th}$ release of
the Sloan Digital Sky Survey (SDSS, \citealt{adelman2008}).
\subsection{{\it INTEGRAL}/OMC data}
\begin{figure}
\centerline{\psfig{figure=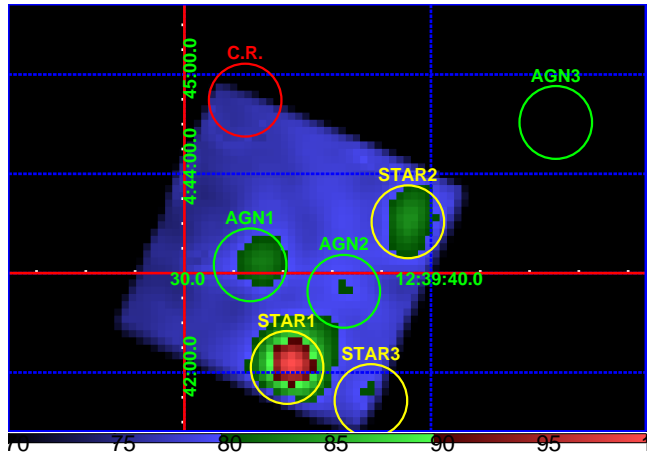,width=85.mm,height=60.mm}}
\caption{Mosaic image obtained with OMC simultaneous to the {\it AGILE} detection of gamma-ray flare of \gbsource. The pixel size of the mosaic image is 1/3 of the original OMC pixel. The label \emph{AGN1} is for \gbsource\ (\sdsssource),
\emph{AGN2} for SDSS J123936.52+044249.3, \emph{AGN3} for SDSS J123945.06+044431.4. The circle marked {\em C.R.} indicates the filtered region where a cosmic-ray event was collected during revolution 635.}
\label{fig:omc_mosaic}
\end{figure}

We analysed the OMC (Optical Monitoring Camera, equipped with a Johnson V filter, see \citealt{mas-hesse2003}) data collected during period A with the OSA software version 8.0. Due to the dithering mode of {\it INTEGRAL} observations, the field around the source was observed with OMC in the science windows 2, 50, 52, 78 of revolution 633, in the science windows 49, 51, 77 of revolution 635, and in the science windows 49 and 51 of revolution 637. To minimize readout noise contribution to the signal to noise ratio (S/NR), we selected snapshots of 200 s.
Typically 6-10 snapshots of at least 200 s integration were recorded for each science window. Due to telemetry limits, the whole frame of the OMC field of view cannot be telemetred, but only sub-frames around sources from a reference catalog. \gbsource\ is included in one of the predefined sub-frames.
We detected the source with S/NR=4 and m$_V$=17.5$\pm$0.3 with an aperture of 3$\times$3 pixels, averaging the 3 revolutions. The investigation of previous {\it INTEGRAL} observations of the field resulted only in upper limits (m$_V$~$>$ 18.1--19.1
depending on the exposure of the observations). 
The Mosaic image from the individual snapshots taken during the multiwavelength campaign (obtained using the {\sc iraf} package) is shown in Fig. \ref{fig:omc_mosaic}. The pixel size of the mosaic image was 1/3 of the OMC pixel to reduce spurious effects in the images shift.
A cosmic ray signal was filtered out (affecting the combined image around $\alpha_{2000}$=12 39 32.5, $\delta_{2000}$=+04 44 44.5), for science window 77 of  revolution 635 only. This region is marked as C.R. in Fig.
\ref{fig:omc_mosaic}.
\subsection{VLA archival data}
We retrieved archival Very Large Array (VLA\footnote{The National Radio Astronomy
Observatory is a facility of the National Science Foundation operated under cooperative
agreement by Associated Universities, Inc.}) D-array 43 GHz data of
\gbsource, observed on 2001 November 5 for 65 s.
The VLA data were reduced and analysed using standard routines implemented in the
Astronomical Image Processing System (AIPS) package. The sources 3C\,286 (1.45 Jy) and
1254+116 (0.43 Jy) were used as flux and phase calibrators, respectively.\\
An almost unresolved source has been detected at 43 GHz with position
$\alpha_{\rm J2000}$=12$^{\rm h}$39$^{\rm m}$32\fs75$\pm$0\fs01 and
$\delta_{\rm J2000}$=04\degr43\arcmin05\farcs2$\pm$0\farcs1,
a peak flux density of $\sim$71.4$\pm$3.6 mJy,
and an integrated flux density of $\sim$81.2$\pm$4.1 mJy.
\section{The archival data}
\begin{figure*}
\centering
\begin{tabular}{c}
\psfig{figure=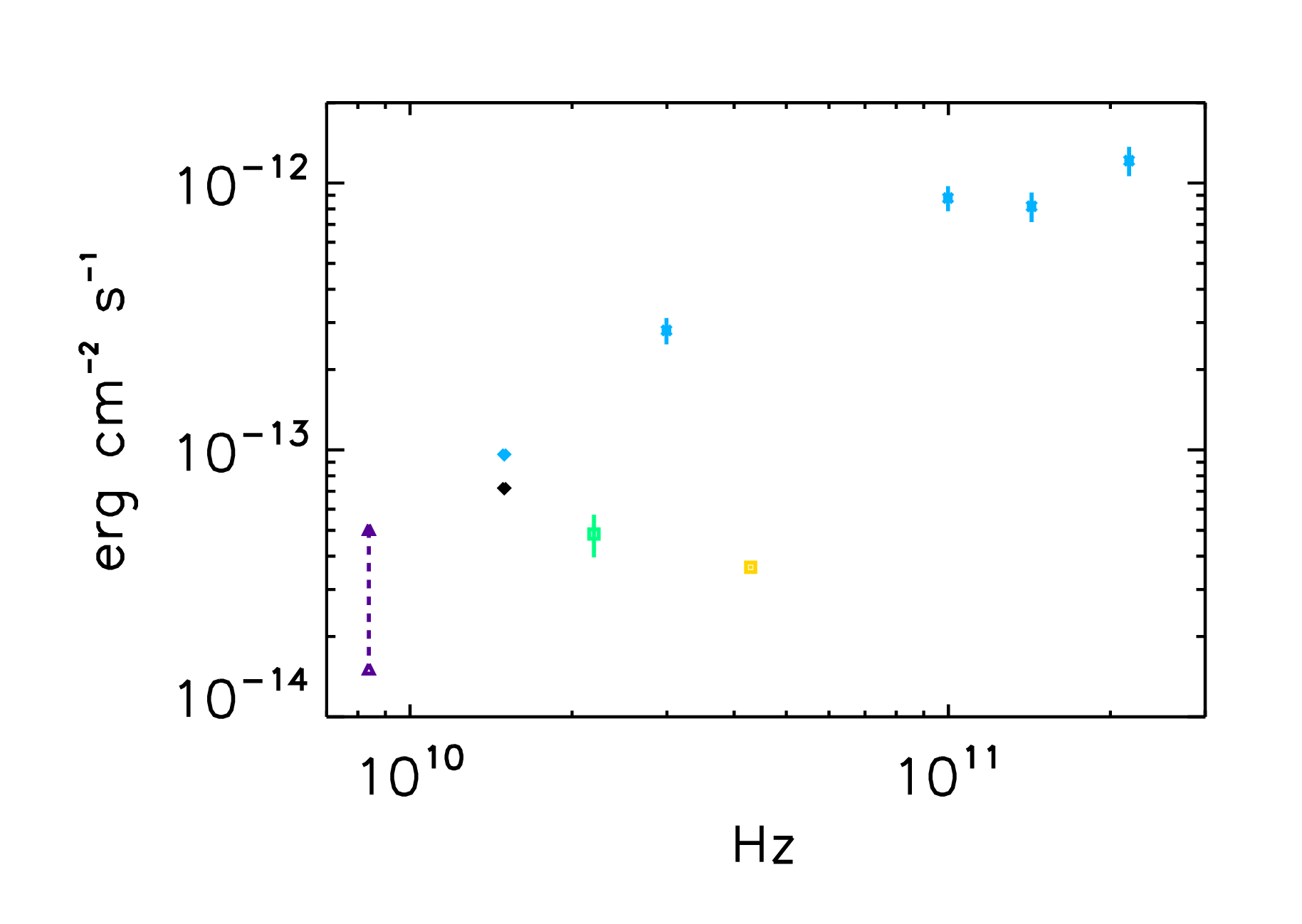,width=130.mm,height=105.0mm} \\
\psfig{figure=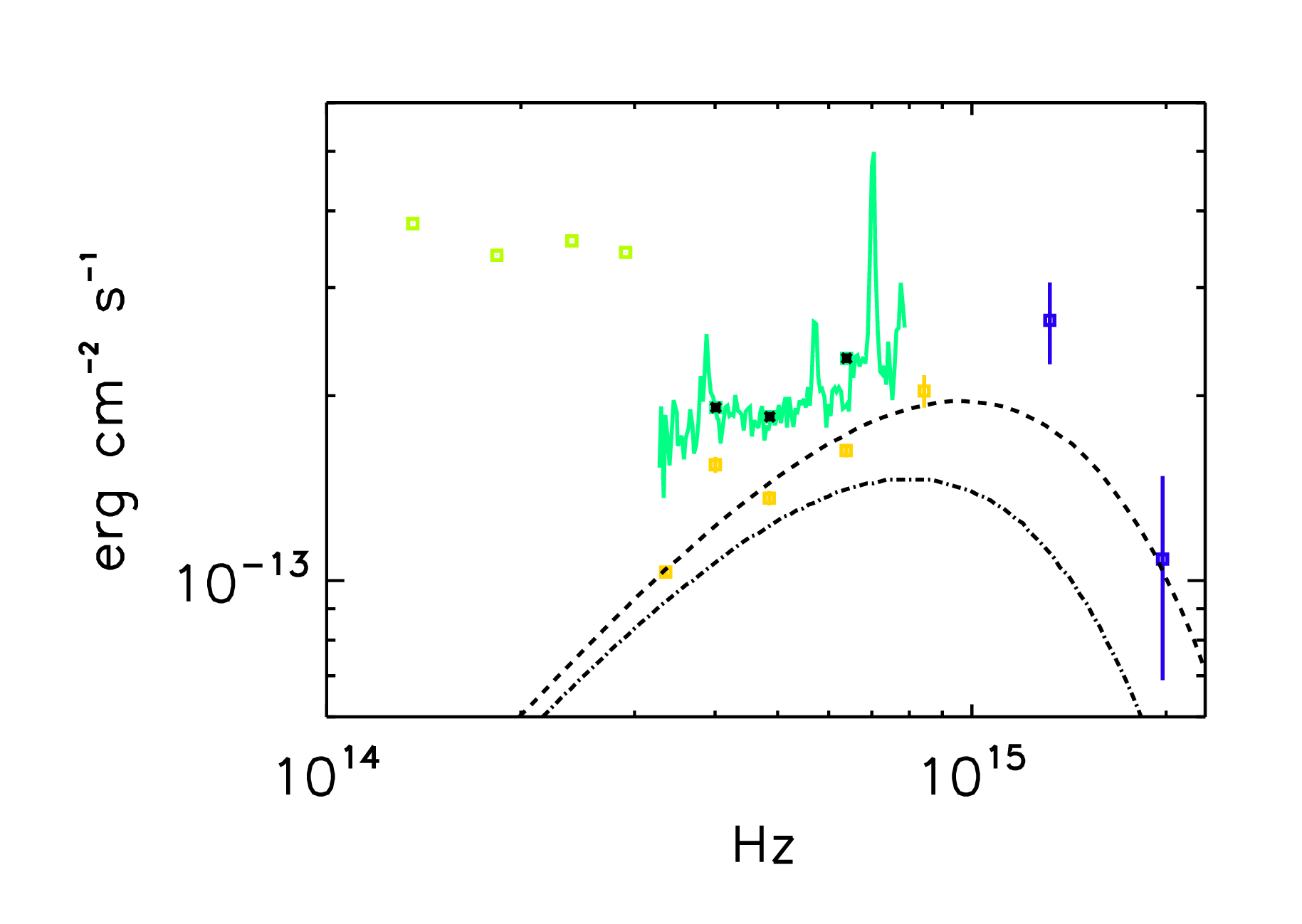,width=130.mm,height=105.0mm}
\end{tabular}

\caption{Top panel: archival radio observations of \gbsource.
  Yellow box data was taken by VLA in 2001, 
  Green data  was taken by Metsahovi in 2002.
  Cyan diamond taken by VLBA in December 2009.  Cyan stars
  are for Planck taken in 2010. Purple data represent the higher and
  lower fluxes measured at 8.4 GHz and reported in literature (from
  Mojave database).  Bottom panel: archival optical observations of
  \gbsource, all data are Galactic absorption corrected. Yellow squares
  are SDSS photometry taken in 2001; the green line is the SDSS optical
  spectrum taken in 2002 (squares of the same color with superposed
  black stars represent the simulated SDSS photometry evaluated from the
  observed spectrum). Light green squares show near-IR photometry taken
  by UKIDSS-LAS at the beginning of 2009, and blue squares are the GALEX
  data taken in mid-2007. The dashed and dot-dashed lines represent two
  disk-emission models; see text for details.}

\label{fig:radio_optical_archive}
\end{figure*}

In order to model accurately the SED (sections 5.2 and 5.4), we used archival data of \gbsource\ from the  SDSS. Photometry with the filters u, g, r, i, z was performed on 2001 March 15.
In particular the SDSS archive reports $m_u$=20.62$\pm$0.06 and $m_g$=20.47$\pm$0.03 (corresponding to $m_U \sim$19.9, and $m_V \sim$20.5).
An optical spectrum (370--920 nm) was obtained on 2002 May 13.\\
The near infrared photometry of the source was performed by using the data of the UKIDSS (UKIRT Infrared Deep Sky Survey) LAS (Large Area Survey) of 2007 January 18.\\
GALEX (an orbiting ultraviolet space telescope, see \citealt{galex2005}) observed the source with NUV and FUV filters (with bandpass centered at 230 and 150 nm respectively) between 2007 April 17 and 2007 May 13.\\
The source was observed by VLA at 43 GHz on 2001 November 5, resulting in a low flux (we reported the analysis of this observation in the previous section).\\ 
A weak detection at 22 GHz was obtained by the Metsahovi observatory on 2002 May 11 \citep{Terasranta2005}, they reported a flux of 0.22$\pm$0.04 Jy.\\
Following the {\it Fermi}--LAT detection of a gamma-ray flare, the source has been
added to the MOJAVE\footnote{MOJAVE (Monitoring Of Jets in Active galactic nuclei with VLBA Experiments) is a long-term program to monitor radio brightness and polarization in jets of AGN.} sample, and was observed twice by VLBA in 2009 at 2 cm: on 2009 January 30 (one month after the gamma-ray flare) and on 2009 December 10.
The source was found in the Planck legacy archive v0.5, with detections at 30, 100, 143, 217 GHz with observations on 2010 January 3, 2010 January 6, 2010 January 9, 2010 January 7, respectively. We also added the 147~GHz data for which the detection is flagged as {\em extended}. At last we included in the SED the detection in the {\it ROSAT} all Sky Survey.
A summary of the archival observations is given in table \ref{tab:archival}, while the archival radio and optical SED is plotted in figure \ref{fig:radio_optical_archive}.
\begin{table}
\caption{Summary of archival optical and radio observations of \gbsource.}
\label{tab:archival}
\centering
\begin{tabular}{c l c}
\hline\hline
date & project/observatory & measurement description \\ \hline
2001 March 15   & SDSS & optical photometry \\
2001 November 5 & VLA & 43 GHz \\
2002 May 11 & Metsahovi & 22 GHz \\
2002 May 13 & SDSS & Optical Spectrum  \\
2007 January 18 & UKIDSS large Area Survey & near infrared photometry \\
2007 April 17 \& May 13 & GALEX & near UV photometry \\
2009 January 30 & VLBA & 15 GHz \\
2009 December 10 & VLBA & 15 GHz \\
2010 January 3--9 & Planck & 30 -- 217 GHz\\
\end{tabular}
\end{table}
\section{Results}
\subsection{Disk luminosity and black hole mass determination from the archival optical/UV photometry}
In spite of the large amount of data available for \gbsource\,
%GG% that we have summarized in the previous section,
the source has not been studied in
detail before.\\
The optical SDSS spectrum was taken simultaneously with the 22 GHz
  Metsahovi radio observation \citep{Terasranta2005}. We note that the 22 GHz flux is a factor
  3--4 lower than the Planck data of 2010 (light blue data in top panel
  of fig.  \ref{fig:radio_optical_archive}). From the optical spectrum
  shape (bottom panel of fig. \ref{fig:radio_optical_archive}) and from
  the faint radio emission, a reasonable assumption is that the optical
  emission in May 2002 is accretion-disk-dominated.  We evaluated the
  non--thermal dominance ({\it NTD}, see \citealt{shaw2012}) of the
  optical spectrum making use of the \ion{C}{IV} line luminosity, and
  the continuum luminosity extrapolated at 1350\AA\ rest-frame.

We obtained {\it NTD}=1.1$^{+6.7}_{-1.0}$ where the major contribution to the errors comes from
the uncertainties in equation (3) of \cite{shaw2012}, and {\it NTD}=1.0$^{+0.8}_{-0.5}$
making use of the \ion{Mg}{II} line luminosity and the continuum luminosity at 3000 \AA\ rest-frame,
indicating that the optical spectrum taken in 2002 is typical of a thermal emission
  dominated source. We also
  observe that the optical photometry taken by SDSS in March 2001
  results in a similar shape and lower flux ($\sim$ 30\% lower), and that the radio
  observation by VLA in November 2001 shows the source in a rather radio
  faint state.  Therefore we also make the assumption of accretion disk
  dominated emission for the optical photometry taken in March 2001.\\  

  We compare the optical data taken by UKIDSS-LAS in January 2007, the
  SDSS spectrum taken in May 2002, and the optical photometry taken in
  March 2001.  These observations taken at different epochs seem to
  demonstrate three different emission states: a jet-dominated emission
  state (UKIDSS-LAS), an accretion-disk-dominated state (SDSS
  photometry), and an intermediate state (SDSS optical spectrum) that
  seems to be dominated by accretion-disk emission with a possible minor
  contribution from the jet.  We also note that the GALEX observations
  give a factor of $\sim$10 lower flux with respect to the {\it
    Swift}/UVOT observation taken during the 2009 January flare.  We
  cannot establish whether the UV emission detected by GALEX is due to
  the disk alone or whether there is also the contribution of the high energy
  tail of synchrotron emission. In the latter, unfavoured, case,
  the UV data from GALEX result in at least an upper limit on the disk emission.\\
  
  With these considerations, and assuming the disk emission is not
  variable on year timescales, we attempted to model the SDSS and GALEX
  photometry with a Shakura Sunyaev accretion disk \citep{shakura1973}
  around a non-rotating black hole (dashed curve in the bottom panel of
  Fig. \ref{fig:radio_optical_archive}).  We modeled the disk emission
as proposed by \cite{ghisellini2009}, with inner radius of 3
Schwarzschild radii ($R_S$), and outer radius of 500 $R_S$.  We fit the
model to the SDSS+GALEX data keeping free the parameters of disk
luminosity ($L_d$) and $R_S$.  We obtained a disk luminosity of
$\sim$8.9$\times 10^{45}$ erg s$^{-1}$, and $R_S$$\sim$2.4$\times
10^{14}$cm, corresponding to a maximum emitting temperature of
$\sim$5.4$\times$10$^4$K, a BH mass of $\sim$8$\times 10^8$ solar masses,
an accretion rate of $\sim$9
M$_{\mbox{\footnotesize{\sun}}} y^{-1}$
(assuming an accretion efficiency $\epsilon_{accr}$=0.1), and then an Eddington ratio R$_{Edd}=L_d/L_{Edd} \sim 50\%$.\\
  
 We note that one observation by SDSS and one observation by GALEX
  show a flux higher than the one expected according to our thermal
  emission model, but the filters used include quasar line emission from
  Mg\,{\sc ii} (the SDSS {\it i} filter) and Ly$\alpha$ (the GALEX NUV
  filter); see Figure \ref{fig:radio_optical_archive}.
  We also fit a disk model assuming the near-UV (GALEX) photometry is
  dominated by jet emission.  In this case, the model assumes an
  accretion disk with only a minor contribution from jet emission and is
  fit to the SDSS data alone.  In this scenario (dot-dashed curve in the
  bottom panel of Fig. \ref{fig:radio_optical_archive}), a model with a
  disk luminosity $>$5.4$\times 10^{45}$ erg s$^{-1}$ is required.  We
  are aware that our model fits are not unique and that there are no
  simultaneous radio observations to strongly validate one fit over the
  other, but our preferred model reproduces the observed fluxes and is
  consistent with the scenario usually proposed for FSRQs (see for
  example \citealt{ghisellinihighz}, where hard optical/UV spectra are
  usually modeled as accretion disk dominated emission).
\subsection{Black Hole Mass Determination from Archival Optical Spectrum}
\begin{figure}
\centering
\psfig{figure=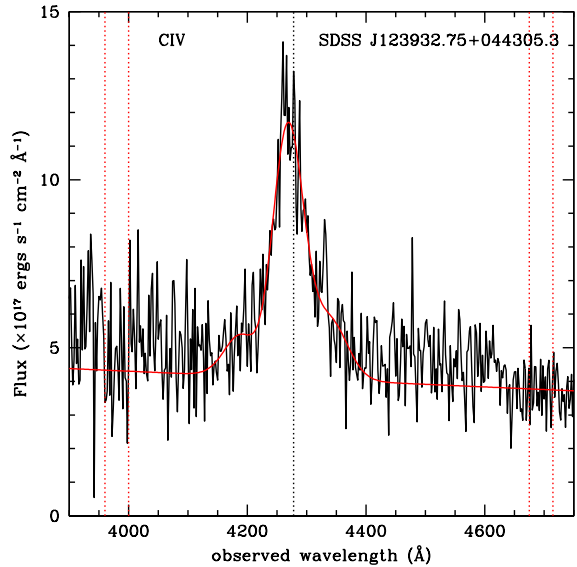,width=80.mm,height=80.mm}
\caption{The  C\,{\sc iv} line in the archive optical spectrum of \gbsource. The red continuum line is the fit with a sixth-order Gauss-Hermite polynomial.}
\label{fig:c4andfit}
\end{figure}

The BH mass of \gbsource\ could be estimated better by using the single-epoch
BH mass scaling relationship for C\,{\sc iv} derived from
\cite{Vestergaard2006} and applied on the archival optical spectrum for this FSRQ.
Unfortunately, the SDSS spectrum has a
rather low S/NR ($\sim3$ per pixel in the continuum near the C\,{\sc
  iv}$\,\lambda$1549 emission line).
\cite{denney2009} show that
line widths and thus single-epoch BH masses measured from low
S/NR data have relatively larger systematic uncertainties than those
measured from high quality data.  In addition, unrecognized absorption
is a particular concern for low quality C\,{\sc iv} data
%GG%(see Vestergaard \& Peterson, 2006; Assef et al.\ 2010; Denney et al.\ 2011).
\citep[see ][]{Vestergaard2006,assef2011,denney2011}.
Nonetheless, this is the only optical spectrum during a low state
currently available.  Therefore we use this spectrum to measure the line width and determine the mass.
We employed two line width measurement
methods and quote conservative uncertainties on the line width taking into
account the low quality of the data.  Our methods for measuring the
C\,{\sc iv} line width and uncertainties closely follow the prescription
``A'' described by \cite{assef2011}, and we therefore refer the
reader to this work for details.  After subtracting the linearly fit
continuum, based on the windows shown in Fig. \ref{fig:c4andfit},
we measured the FWHM of C\,{\sc iv} both directly from the data (${\rm
  FWHM}=2860\pm910 \ {\rm km s}^{-1}$) and from a sixth-order
Gauss-Hermite polynomial fit to the line profile, as shown in Fig.
\ref{fig:c4andfit} (${\rm FWHM}=4710\pm390 \ {\rm km s}^{-1}$).
\cite{denney2009} show that direct measurement of the FWHM from low S/NR
data systematically underestimates the line width, while measurement
from a fit referring to the same data can overestimate the same width.
We adopt a conservative approach and take the mean of these two width
measurements and assume the quadrature sum of the uncertainties.  The
adopted FWHM measurement becomes ${\rm FWHM_{CIV}}=3800 \pm 1000$ km
s$^{-1}$.  We then measure the mean continuum luminosity in the
continuum window near rest-frame 1450\AA\ to be $\lambda
L_{1450}=(3.47\pm0.44)\times 10^{45}$ erg s$^{-1}$, after correcting
for Galactic extinction.  We then evaluate the mass using equation (7)
from \cite{Vestergaard2006}, which is also Equation (6) from
\cite{assef2011}.  It is worth noting that the SDSS spectrum does not
extend to rest-frame 1350\AA; however, \cite{Vestergaard2006} argue that
the 1450\AA\ luminosity can be substituted without penalty for the
1350\AA\ luminosity, as we have done here.  We estimate the BH mass of
\gbsource\ to be
$4.3^{+3.2}_{-2.2}\times 10^8 {\rm M}_{\mbox{\footnotesize{\sun}}}$.\\
%GG%
\cite{assef2011} find a correlation between the ratio of the C\,{\sc iv}-to-Balmer mass estimates
and the UV-to-optical luminosity ratio.  Since this
correlation is based on the {\it ratio} of the mass estimates, barring further
investigation into the source of this correlation,
it is unclear whether
it is the C\,{\sc iv}-based or Balmer-based mass estimates, or both, to be
the source of the bias.
Regardless of origin, Assef et al.\ found that,
when this correlation is removed, and when they arbitrarily choose to
correct the C\,{\sc iv}-based masses, the corrected masses are
highly consistent with the measured Balmer-line-based mass estimates (the
scatter in the corrected C\,{\sc iv} vs.  Balmer mass estimates is reduced by a
factor
%GG%$\gtrsim$ 2
$\ga$ 2
when compared to the uncorrected mass estimates; see \citealt{assef2011}). 
Balmer-based mass estimates are generally more accepted in
the literature because they are relatively better calibrated with direct
mass measurements from reverberation mapping
\citep[see, e.g., ][]{collin2006,Vestergaard2006,denney2009}.  At this point, however,
it is impossible to state which of the two mass measures is
actually more accurate.  In particular, we must consider that host
galaxy starlight can significantly contaminate the optical luminosity
with which Balmer-based masses are estimated, yet there may be evidence
of non-virial motions from C\,{\sc iv}
%GG%(Richards et al.\ 2002, AJ, 124, 1). 
\citep[see, e.g., ][]{richards2002}.
For \gbsource, we fit a powerlaw continuum to
the full wavelength extent of the SDSS spectrum and extrapolate to
rest-frame 5100\AA\ to estimate the rest-frame optical luminosity to be
$\lambda L_{5100}=(2.64\pm0.33)\times 10^{45}$ erg s$^{-1}$.  Using
equation (8) of \cite{assef2011}, with the coefficients based on their
prescription A, we can then calculate a corrected C\,{\sc
  iv}-based BH mass of
$5.3^{+4.4}_{-3.3}\times 10^8 {\rm M}_{\mbox{\footnotesize{\sun}}}$.\\
The two mass estimates (based on the C\,{\sc iv} broad
line width and on the thermal continuum) are in qualitative agreement.
\subsection{Gamma-ray light curve}
\begin{figure*}
\centering
\psfig{figure=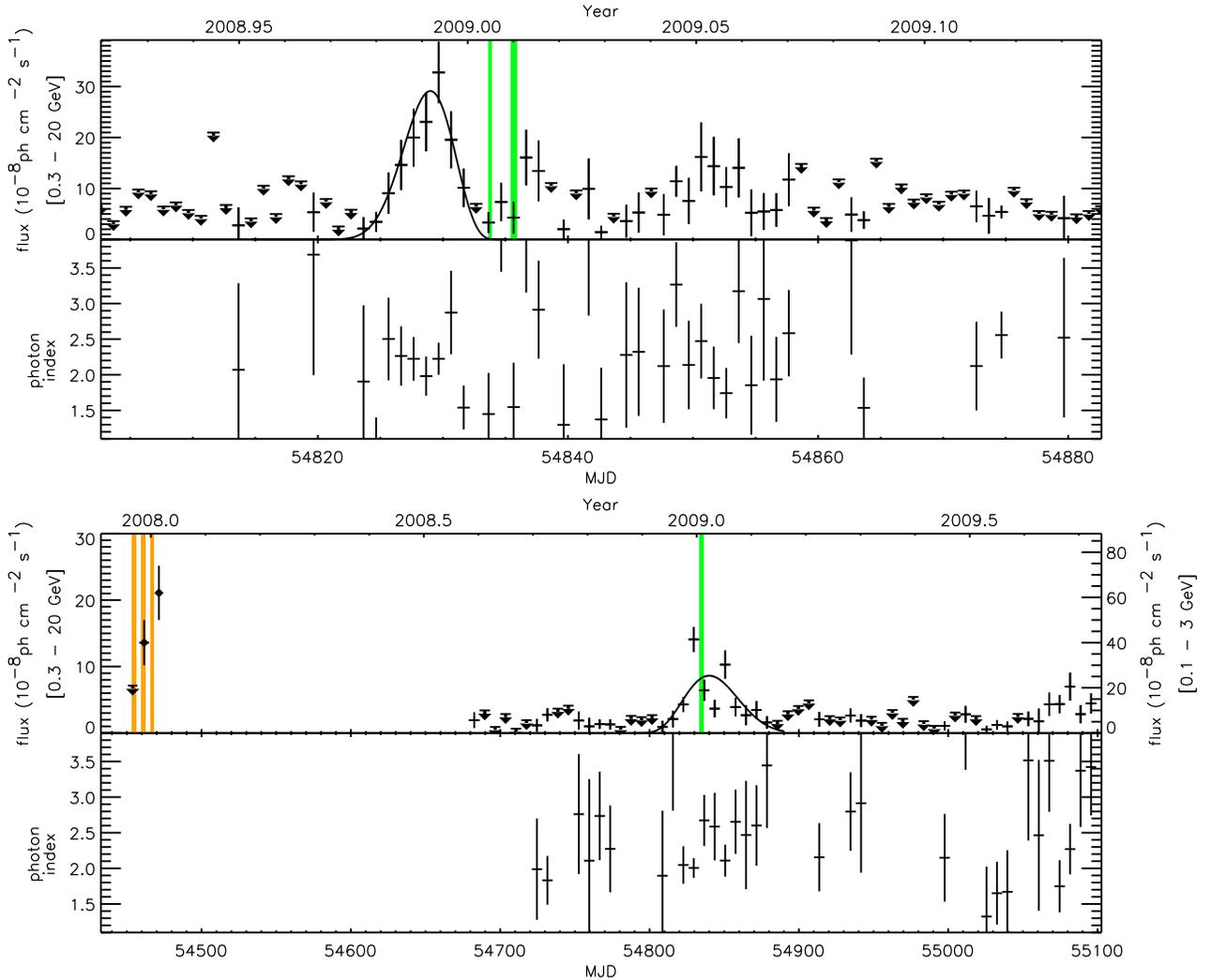,width=170.mm,height=140.mm}
\caption{The gamma-ray light curve for \gbsource\ obtained with {\it AGILE} during campaign A (diamond symbols and right vertical scale for the gamma-ray flux) and {\it Fermi} (left vertical scale for the gamma-ray flux).
For the {\it Fermi}--LAT data we report also the photon index evaluated in the 300 MeV - 20 GeV energy band. The upper panel
reports {\it Fermi}--LAT data with a typical binsize of 1 day. The lower panel reports {\it Fermi}--LAT data with 7 days integration. {\it AGILE} data were integrated with bin sizes of about 6.5, 6.5, 4 days, due to gaps in the observation
(see Table \ref{tab:obs}).
The orange bands represent the {\it INTEGRAL} campaign (campaign A); width is in scale. The green bands represent the {\it Swift} campaign (campaign B); width not in scale. }
\label{fig:lc}
\end{figure*}
We created a gamma-ray light curve for the source from both the {\it AGILE} pointing and the {\it Fermi} survey, as shown in Fig. \ref{fig:lc}.
The {\it AGILE} data were integrated with bin sizes of about 6.5, 6.5, and 4 days, due to gaps in the observation (see Table \ref{tab:obs}). The {\it Fermi}--LAT data
were integrated with binsizes of 1 and 7 days.
The flux reported is in the 100 MeV - 3 GeV range for {\it AGILE}.
As discussed in section 3, the {\it Fermi}--LAT data were analysed between 300 MeV and 20 GeV in order to reduce the contamination from the nearby 3C 273 at lower energies.
In Fig. \ref{fig:lc} we also report the gamma-ray photon index in the 300 MeV - 20 GeV as obtained by the {\it Fermi}--LAT data assuming a power-law spectrum.\\
To evaluate the flare duration, we made use of the procedure
  described by \cite{abdo3c273}. From the light curve with time bin of 1
  day, we obtained a duration (defined as $\frac{T_{rise}+T_{fall}}{2}$) of 7 days
  and an asymmetry -0.3 (see \citealt{abdo3c273} for details, $T_{rise}$ and $T_{fall}$ are the rising and falling time respectively).  The
  light-curve with time bin of 7 days shows more than a single relative
  maximum; therefore, the fit with a simple curve is not feasible.  A
  rough definition of the gamma-ray activity period could instead be the
  timespan for which {\it Fermi}--LAT detects gamma-ray emission from the source.
  Assuming temporal bins of 7, 15, and 30 days, we found that {\it Fermi}--LAT
  detected gamma-rays from the source for at least 11 weeks.
\subsection{The {\it Fermi}--LAT gamma-ray spectrum for the high activity period of campaign B}

\begin{figure}
\centering
\psfig{figure=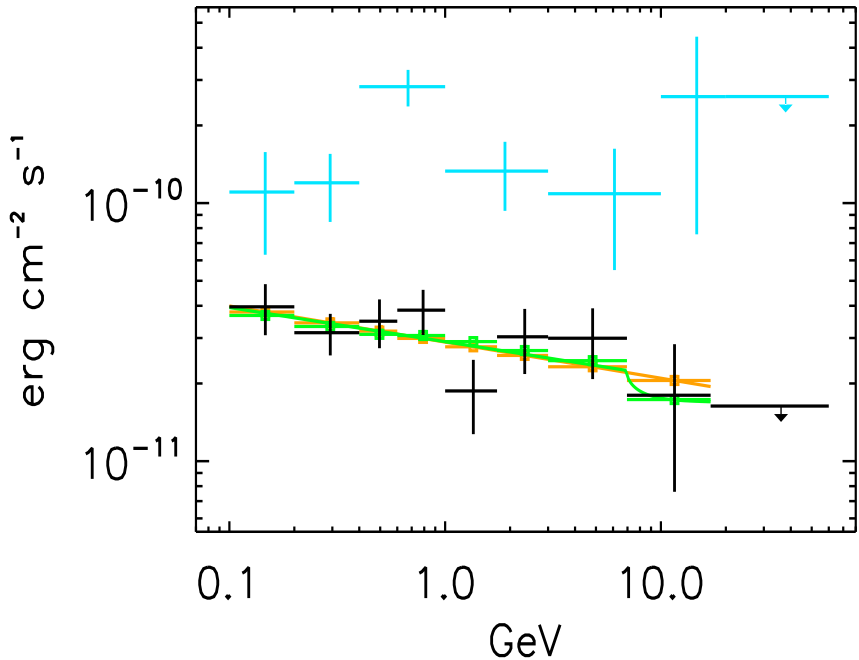,width=130.mm,height=105.mm}
\caption{Gamma-ray spectrum of \gbsource\ integrated for one month around the peak flux (black data). The spectrum integrated for 4 days around the flare is reported too, multiplied by a factor 2 (cyan data).
The fitted models discussed in the paper are reported. The continuous lines are for the models, the binned curve are for models weighted over each energy bin.
The power law model is reported in orange, the power law model + gamma-gamma absorption is reported in green.}
\label{fig:gammaspectrum}
\end{figure}

The gamma-ray spectrum obtained by {\it Fermi}--LAT data, integrated for
  one month centered around the peak flux (2008 December 29 16:00 UT),
  is reported in Figure \ref{fig:gammaspectrum} together with the
  spectrum integrated for four days.
The 30-day integrated spectrum has been fit first with a power-law
model.  Because absorption is expected for a blazar-zone originating near
the central source (e.g., within the BLR), we also fit the spectrum with a
power-law combined with absorption.  In particular, absorption
at E$>$ 20 GeV/(1 + z) is expected \citep{tavecchio2009} due to the Ly continuum, or at E$>$5 GeV/(1 + z) due to the He\,{\sc ii} recombination continuum
\citep{poutanen2010}.  We therefore fit the
gamma-ray spectrum of \gbsource\ with a power-law combined with the
gamma-gamma absorption model as proposed by \cite{poutanen2010}.
Here, the absorption was fit with two parameters: (1) the optical depth for the H\,{\sc i} complex
 ($\tau_{H}$) and (2) the optical depth for the
He\,{\sc ii} complex ($\tau_{He}$).
We fit the models to the data for gamma-ray energies below 20 GeV, because the extragalactic background light is expected to
absorb gamma-rays of E$\ga$20 GeV, for sources at the redshift of \gbsource\ \citep{finke2010}. 
The results of the fit are reported in table \ref{tab:sfit}.
\begin{table}
\caption{Gamma-ray spectral properties of \gbsource, uncertainties at 68\% C.L.}
\label{tab:sfit}
\centering
\begin{tabular}{c c c c c c c}
\hline\hline
\multicolumn{2}{c}{Power Law} & & \multicolumn{4}{c}{Power Law + Double Absorber} \\ \cline{1-2} \cline{4-7} 
$\chi^2/DoF$ & photon index & & $\chi^2/DoF$ & photon index & $\tau_{H}$ & $\tau_{He}$ \\ \hline
1.1  &  2.14$\pm$ 0.08 &  & 1.6  & 2.13$\pm$ 0.08 & 1.0$^{+4.6}_{-1.0}$ & 0$^{+0.9}_{-0}$  \\ \hline
\end{tabular}
\end{table}
The fit with the gamma-gamma absorption components results in no
absorption from the He\,{\sc ii} complex, and weak absorption from
the H\,{\sc i} complex.
We performed the F--test on the two fits reported in table \ref{tab:sfit}
to test the hypothesis of the need of the absorption components. The
F--test gives a value of 0.15, and an associated probability of
$\sim$87\%, hence the absorption component is not necessary to fit the
spectrum, suggesting that in
\gbsource\ the blazar zone is located in the outer low-ionization region of the BLR or outside it.
\subsection{The spectral energy distribution for the high activity periods}

\begin{figure*}
\centering
\begin{tabular}{c}
  \psfig{figure=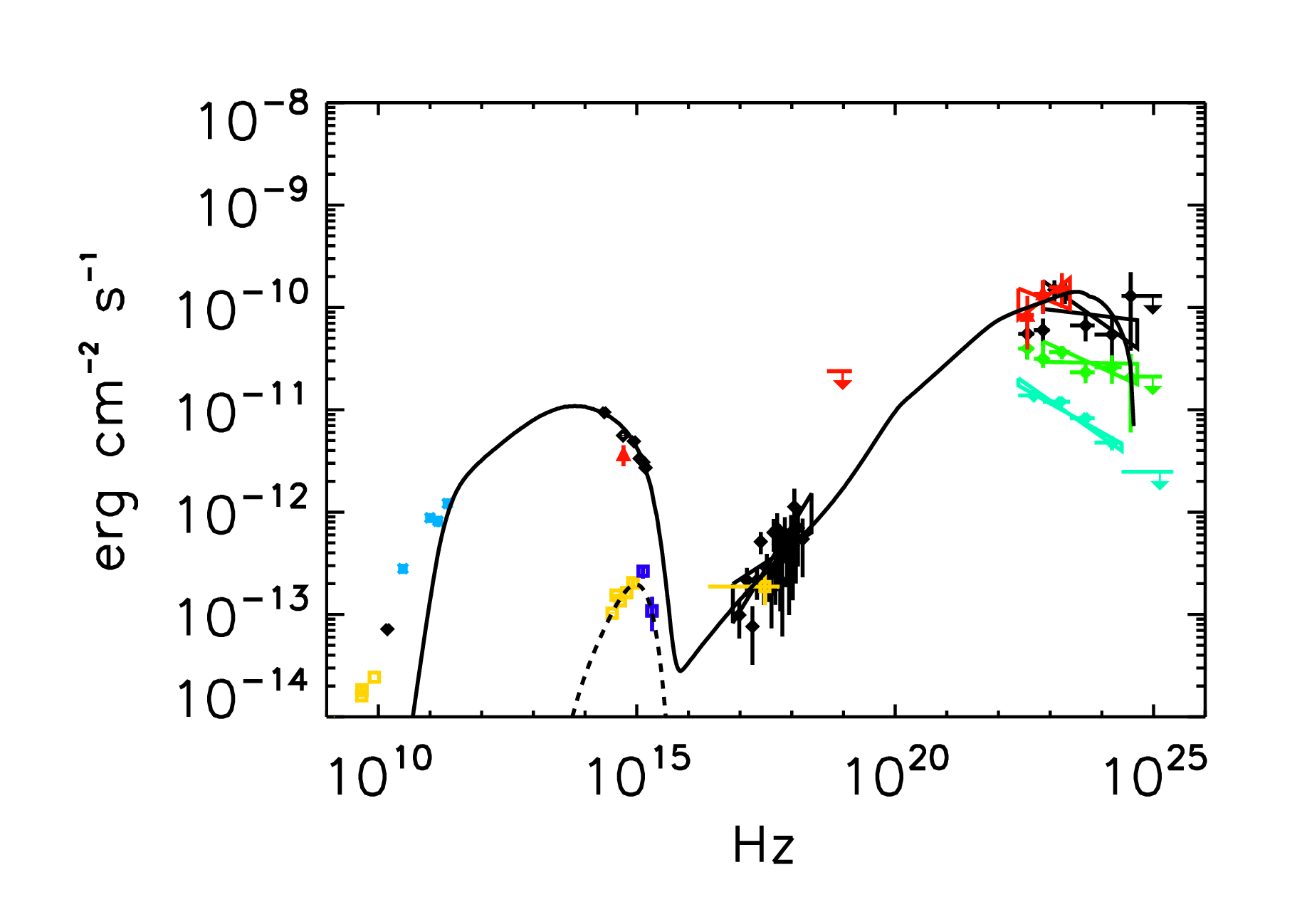,width=130.mm,height=105.mm} \\
  \psfig{figure=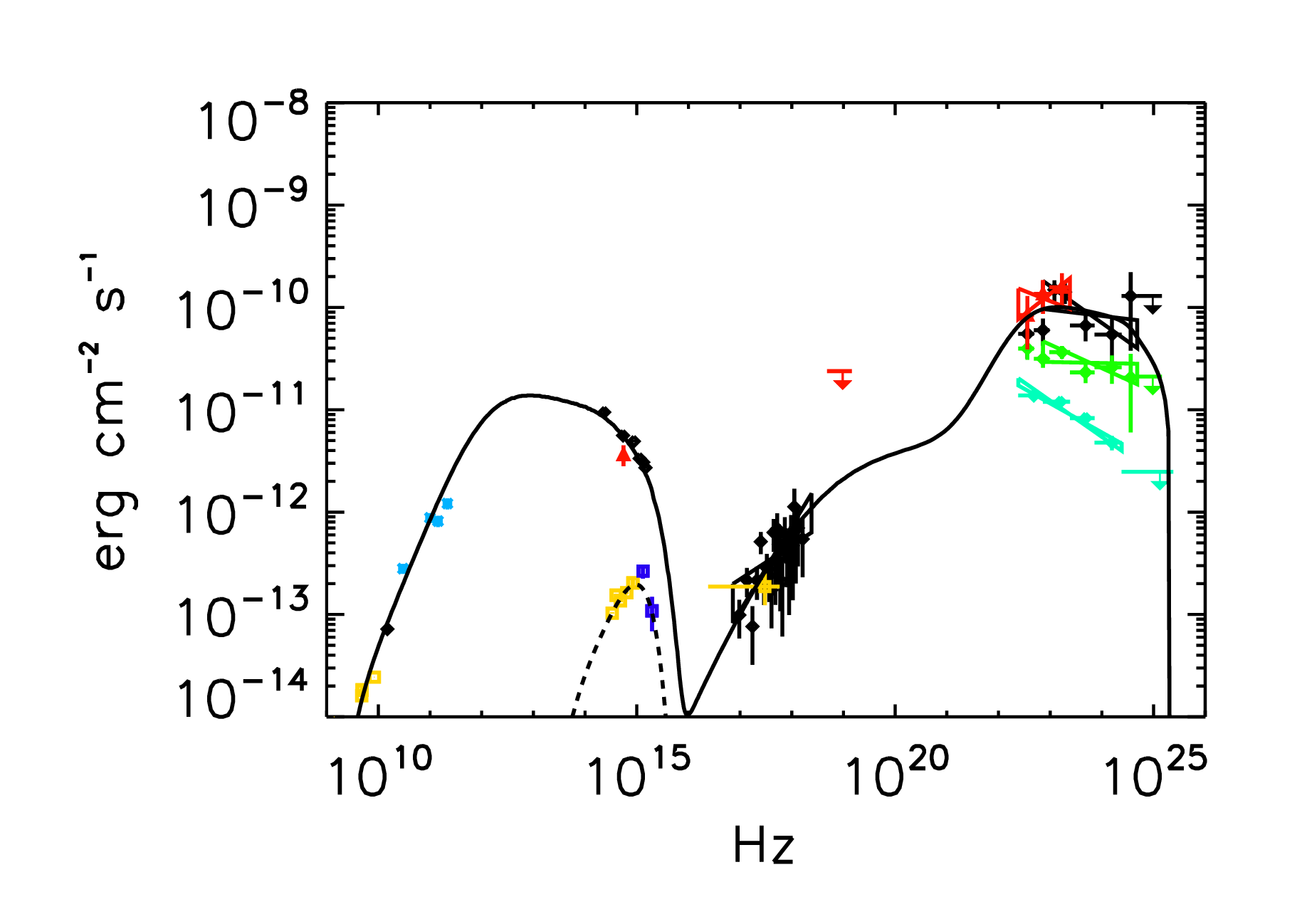,width=130.mm,height=105.mm}
\end{tabular}
\caption{In the two plots we report the SEDs for the two high gamma-ray activity periods detected with {\it AGILE} (red symbols, data from {\it INTEGRAL}/OMC, {\it INTEGRAL}/IBIS, {\it AGILE}--GRID)
and {\it Fermi}--LAT (black symbols, data from VLBA, Kanata, {\it Swift}/UVOT, {\it Swift}/XRT, {\it Fermi}--LAT) respectively.
Green symbols are for data from {\it Fermi}--LAT integrated for one month period centered on the gamma-ray flare; light-cyan symbols are from the second {\it Fermi}--LAT catalog.
We report optical data from SDSS (yellow)
and UV data from GALEX (blue), as well as the Planck data (cyan), and the {\it ROSAT} detection (yellow) during the all-sky survey. Archival radio data from NED are shown in yellow.
The disk model for the low activity period is plotted as a black dashed line; the model for jet emission in high activity period is reported as a continuous black line.
In the top (bottom) plot we show the model assuming a blob dissipating at 0.2 pc (7 pc) from the central BH.}
\label{fig:sed_multiepochs}
\end{figure*}

We constructed two spectral energy distributions referring to the high
gamma-ray activity observed by {\it AGILE} and {\it Fermi} (campaigns A
and B, respectively, as reported in Table \ref{tab:obs}) with the data
collected so far.  The {\it AGILE} data are integrated for 4 days (the
duration of the last observation during the campaign A, from 2008
January 4 13:35 to 2008 January 8 11:06 UT).  The {\it Fermi}--LAT data for
period B are integrated within 4 days around the flux peak (2008
December 29 16:00 UT) to achieve acceptable statistics.
%GG%these data are simultaneous to within a few days.
These spectral energy distributions
are shown in Figure \ref{fig:sed_multiepochs}, where we also included
the archival data.\\
\begin{figure*}
\centering
\begin{tabular}{c}
  \psfig{figure=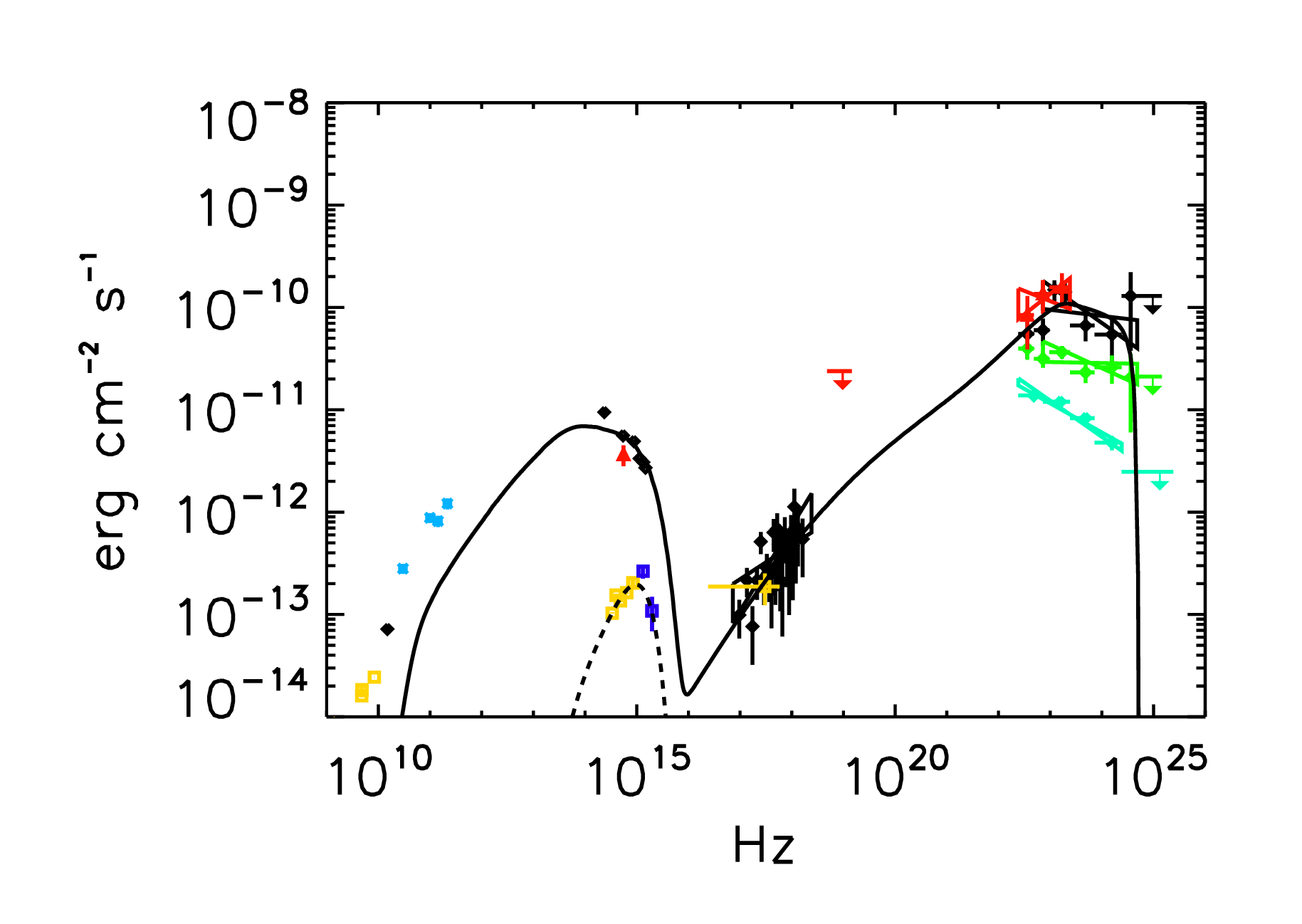,width=130.mm,height=105.mm}
\end{tabular}
\caption{Same as previous figure, but in the jet emission model we assume a blob dissipating at $\sim$5 pc, relaxing the condition
R$_{blob}$=$\frac{1}{10}$R$_{diss}$, and assuming a blob radius of $\sim~10^{17}$cm.}
\label{fig:sed_multiepochs2}
\end{figure*}
The SED modeling has been performed in the framework of leptonic models.
We assume that the emitting region is a spherical blob of radius $R_{blob}$, with bulk Lorentz factor $\Gamma_{bulk}$,
an electron population distribution proportional to $\frac{(\gamma/\gamma_{b})^{-s1}}{1+(\gamma/\gamma_{b})^{-s1+s2}}$
(where $\gamma$ is the Lorentz factor of the electrons, ranging from $\gamma_{min}$ to $\gamma_{max}$),
and a randomly oriented magnetic field  $B$ filling the dissipation region. The observer line of sight and the jet direction form an angle $\Theta_{view}$.
We parametrize the external radiation fields as proposed by
\cite{ghisellini2009}, where the key elements are the accretion disk
luminosity and the distance ($R_{diss}$) of the emitting blob from the central BH.
We consider a jet aperture $\frac{R_{blob}}{R_{diss}}$=0.1, according to \cite{ghisellini2009}.
In evaluating the power carried by protons, we assume one proton per emitting electron.
We assumed that during the campaigns A and B, the accretion disk luminosity is the
same as the one obtained from the SDSS data in 2001 (dashed curve in Fig.
\ref{fig:sed_multiepochs}).  We found in section 5.4 that the {\it Fermi}--LAT
gamma-ray spectrum (Fig. \ref{fig:sed_multiepochs} green data, one month
integration) does not show absorption from the BLR.  Therefore, we tried to
find solutions for the blob dissipating beyond the BLR, without taking
into account the gamma-gamma absorption from the BLR.  We obtained two
possible solutions for the modeling.
Model parameters are reported in Table \ref{tab:models},
where we use $R_{BLR}$ ($R_{Torus}$) the refer to the distance of the BLR region (of the dusty torus) from the BH; 
f$_{BLR}$ (f$_{Torus}$) to the fraction of the disk emission that is reprocessed by the BLR in lines  (by the dusty torus);
$\epsilon_{accr}$ to the accretion efficiency; $\gamma_{cooling}$ to the electron Lorentz factor for which the electron energy halves in
the blob crossing time ($R_{blob}$/{\it c}).\\
The first solution (model 1, top panel of Figure
\ref{fig:sed_multiepochs}) places the
dissipation region (R$_{diss}$) just beyond the BLR (R$_{diss}\sim$0.2
pc from the SMBH), where the photon field from the dusty torus is almost
equal to the
BLR seed photon field.  At such a distance, the variability
time-scale of a blob of radius R$_{blob}\sim$7$\times 10^{16}$ cm is of
the order $\sim$3 days, qualitatively in agreement (within a factor 2) with the 1-day binned gamma-ray light
curve of the source.  According to this modeling of the seed photon fields,
the BLR photon contribution is expected to fade over time while the
torus contribution remains constant.
Both SSC and external Compton (EC)
with seed photons from the torus contribute to the soft X-ray  emission, the hard X-ray is dominated by EC emission
with seed photons from the torus, and the GeV emission is dominated by
EC with seed photons from the BLR.  In the MeV up to GeV range
the two EC contributions are almost equal.
Further investigation of the
variability is not possible, as no optical light-curve is available from
which to trace the electron population from the synchrotron emission.
Moreover the evolution of the gamma-ray photon index with time (reported in Fig. \ref{fig:lc}) is statistically poor.\\
The second solution to the SED modeling (model 2) has been found by assuming R$_{diss}\sim$7 pc from the SMBH. Such a model requires a R$_{blob}$ = $2\times 10^{18}$ cm, with a variability time-scale of the order of $\sim$10$^2$~days.
We do not find such a variability time-scale in the gamma-ray light curve, but we observe that the activity period for the source lasts at least 11 weeks. We can argue that the dissipating region in the jet
has a radius of $2\times 10^{18}$ cm. Gradients of the electron energy density in the distance from the black hole, or disturbances in the jet (see for examples \citealt{giannios2010,bromberg2009}) can be responsible for the shorter variability observed in gamma-rays. With this model, the X-ray emission is due to SSC; the gamma-ray emission is due to EC with seed photons from the torus. The direct contribution of accretion disk photons to the EC is negligible.\\
In the model for a blob dissipating just beyond the BLR, the energy region below $\sim$10$^{11}$Hz is self absorbed. On the contrary, in the model of a distant dissipation region,
there is no self-absorption, thus allowing the direct observation of radio emission from the blob. In this case, the emission predicted by the model is slightly higher than the flux reported in Fig. \ref{fig:sed_multiepochs} (bottom panel),
but we note that no radio observations were performed during the gamma-ray flares.
The mismatch in the variability time-scale for model 2 with  R$_{diss}\sim$7 pc, results from the assumption that R$_{blob}$=$\frac{1}{10}$R$_{diss}$. Alternatively, we tried to model the SED assuming a blob of
radius $\sim 10^{17}$cm to satisfy the observed variability time-scale, and dissipating at $\sim$5 pc from the SMBH (model 3).
Model parameters are reported in the right column of Table \ref{tab:models}, and the model is plotted in Fig. \ref{fig:sed_multiepochs2}.\\
A similar kind of solution has been suggested by \cite{tavecchio2011} to explain
faster variability (of the order of 10 min) for PKS 1222+216.
This model can correspond to a reconnection event, causing the acceleration  of a small portion of the jet \citep{giannios2010}, or to recollimation by the interaction with the external medium \citep{bromberg2009}.\\
According to this SED modeling, the X-ray emission is due to SSC, and the gamma-ray emission to EC with external photons from the torus. 

\begin{table}
\caption{Model parameters for the fits of the spectral energy density, in the two assumptions of a blob dissipating just beyond the BLR (R$_{diss}\sim$ 0.2 pc) or far away from the SMBH (R$_{diss}\sim$ 7 pc).
In the last column we report the model parameters for a blob dissipating far away from the SMBH (R$_{diss}\sim$ 5 pc), but relaxing the condition
R$_{blob}$=$\frac{1}{10}$R$_{diss}$, and assuming a blob radius of $\sim~10^{17}$cm.}
\label{tab:models}
\centering
\begin{tabular}{l | c | c | c}
\hline\hline
\footnotesize
                                          & model 1            & model 2                  & model 3 \\ \hline \hline                     
R$_{diss} (pc)$                            & 0.22               & 6.8                      & 4.8\\ \hline  
Blob radius   (cm)                        & 6.7$\times 10^{16}$   $^*$  & 2.1$\times 10^{18}$ $^*$            & 1$\times 10^{17}$ \\ \hline
m$_{BH}$  (m$_{\mbox{\footnotesize{\sun}}}$)   &  \multicolumn{3}{c}{5.3$\times 10^8$} \\ \hline
L$_{d}$ (erg/s)                            & \multicolumn{3}{c}{8.8$\times 10^{45}$}  \\ \hline      
R$_{BLR}$ (cm)                             &  \multicolumn{3}{c}{3.0$\times 10^{17}$} \\ \hline
R$_{Torus} (cm$)                           & \multicolumn{3}{c}{7.4$\times$10$^{18}$}  \\ \hline
f$_{BLR}$                                  &   \multicolumn{3}{c}{0.1} \\ \hline
f$_{torus}$                                &   \multicolumn{3}{c}{0.3} \\ \hline
$\epsilon_{accr}$                          &   \multicolumn{3}{c}{0.1} \\ \hline   \hline
$\Gamma_{bulk}$                            & 20                  & 20        & 20\\ \hline
angle of view (deg)                       & 2                   & 2  & 2 \\ \hline
$\gamma_{min}$                             & 1                   & 1    & 1 \\ \hline
$\gamma_{max}$                             & 3.9$\times 10^3$    & 3.4$\times 10^4$    & 1.3$\times 10^4$ \\ \hline
$\gamma_{break}$                           & 0.95$\times 10^3$   & 1$\times 10^3$      &  1$\times 10^3$ \\ \hline
density at $\gamma_{break}$                &  3.0$\times 10^{-2}$ & 1.5$\times 10^{-4}$  & 9.6$\times 10^{-3}$\\ 
(cm$^{-3}$)\\ \hline                      
s$_1$                                     & 1.1 &      0.5             & 1.3 \\ \hline
s$_2$                                     & 3.1 &      3.3             & 2.5 \\ \hline
B     (Gauss)                             &    6.1$\times 10^{-1}$ &      1.1$\times 10^{-2}$           & 7.6$\times 10^{-2}$ \\ \hline  \hline 
$\gamma_{cooling}$                         & 60         & 1.1$\times 10^4$      & 2.4$\times 10^3$ \\ \hline \hline
electron power                            & 2.2$\times 10^{45}$   &  4.5$\times 10^{46}$  & 1.1$\times 10^{46}$ \\ 
(erg/s)     \\ \hline                     
magnetic power                            & 2.5$\times 10^{45}$   &  7.9$\times 10^{44}$  & 8.6$\times 10^{43}$  \\
(erg/s)     \\ \hline                     
proton power $^{**}$                       & 1.1$\times 10^{47}$   &  1.6$\times 10^{47}$  & 3.0$\times 10^{47}$ \\
(erg/s)     \\ \hline                     
radiated power                            & 3.1$\times 10^{45}$   &  2.5$\times 10^{45}$  & 2.1$\times 10^{45}$ \\
(erg/s)     \\ \hline 
\normalsize
%\hline
%
\end{tabular}
\\$^*$ R$_{blob}$=$\frac{1}{10}$R$_{diss}$ in this model.
\\$^{**}$ We assume one proton per emitting electron.
\end{table}

\section{Discussion and Conclusions}

We rarely have the opportunity to detect the disk emission in FSRQs,
which are generally overwhelmed by synchrotron jet emission \citep[see
][ for example]{pian1999}.  However, our study here suggests
  that we have detected accretion-disk-dominated emission in \gbsource.
  Granted, we cannot fully exclude the possibility that the
  archival SDSS and GALEX observations we have reported could be
  interpreted as other emission mechanisms than thermal disk emission
  because we lack strictly simultaneous radio observations and extended
  radio light curves to corroborate the assumption of low jet emission.
  However the assumption of disk emission remains, in our opinion, the most likely explanation for the observations.\\
The optical observations of \gbsource\ revealed an optical flux
enhancement of a factor 15--30 in 6 years, signifying a shift from
accretion-disk to synchrotron, jet-dominated emission.  The optical
spectrum obtained in the period of faint optical emission allowed the
classification of the source as a FSRQ in BZCAT. We made an estimate of the SMBH
mass of \gbsource\ and the accretion rate from a period of low jet
activity.  With these estimates, we were able to study the December 2008
flare.  Modeling the observed flat gamma-ray spectrum and SED also
allowed for an investigation into the location of the blazar-zone of the
object.\\
As a final remark, it is worth stressing two major points. First, by
definition, our estimate of R$_{diss}$ is model dependent.  The
location of the dissipation region was estimated assuming the parametrization proposed
by \cite{ghisellini2009} for the BLR and the torus contribution to seed
photons for the EC. According to this parametrization, it was possible
to derive R$_{diss}$ from the luminosity ratio of synchrotron to EC
emission. In fact, in the parametrization by \cite{ghisellini2009} for a
dissipation region outside the BLR, the seed photons for EC fade with
distance from the SMBH.  We obtained two solutions. Referring to
Figure~2 of \cite{ghisellini2009}, the ratio
$\frac{U'_{B}}{(U'_{BLR}+U'_{IR})}$ equals the ratio of the optical to gamma-rays
luminosity at the two values of the R$_{diss}$ (assuming knowledge of
the magnetic energy density).
For the flares of \gbsource\ reported in
this article, one solution (model 2) places the blob at R$_{diss}\sim$7 pc, with only the dusty
torus as the origin of seed photons. The other solution (model 1, with
R$_{diss}\sim$0.2 pc) corresponds to just outside the BLR, where the
contribution of seed photons from both the dusty torus and the BLR are
relevant.  The magnetic field is constrained in the SED modeling by the
cutoff of synchrotron emission in the UV (due to the last and most
energetic electrons), and by the corresponding cutoff of EC which we
cannot derive directly from data (that give non-constraining upper
limits at E$>$20 GeV, see Fig.  \ref{fig:sed_multiepochs}).  Assuming
the Thomson regime, and with only one external photon field contributing
to the EC, the ratio {\it f}$_{cutoff}$ between the synchrotron cutoff energy
and the EC cutoff energy is proportional to $\frac{B}{\Gamma_{bulk} <\nu_{seed}>}$,
where $<\nu_{seed}>$ is the typical seed photon field energy.
Hence if
we have constraining data at the highest energy, and with a specific
geometry in the model, we can constrain B/$\Gamma_{bulk}$.
However, the geometry in the model also constrains $U'_{BLR}+U'_{IR}$,
hence in the ideal case we can obtain B/$\Gamma_{bulk}$ and R$_{diss}$
from the SED modeling.
The ratio of synchrotron to SSC luminosity  further constrains the model parameters
$R_{blob}, \Gamma_{bulk}, B$, allowing one to remove the degeneracy between $\Gamma_{bulk}$ and B.
In reality, we can obtain only upper limits of
{\it f}$_{cutoff}$ because the data at higher energies are non-constraining.
So we have only upper limits on $B/\Gamma_{bulk}$ for each
model. As a consequence, the $U'_{BLR}+U'_{IR}$ could be lower than in
our parametrization (we have to maintain the
ratio  $\frac{U'_{B} }{ U'_{BLR}+U'_{IR} }$ at the desired value). This implies that R$_{diss}$ could be higher than our evaluations.\\
The second point is that the photon field intensity is proportional to the accretion disk luminosity, and the BLR and torus location is proportional to $\sqrt L_{d} $.
In all our estimations, we assume that the disk luminosity is almost steady over time, e.g., in the low state observed during the Sloan Survey in 2002, during the GALEX observation in 2007, and during the
gamma-ray flares observed by {\it AGILE} at the beginning of 2008 and by {\it Fermi}--LAT at the end of 2008. This assumption could be false. In this case, we observe that the parametrization of $U'_{BLR}$ and $U'_{IR}$
reported in equation (20) by \cite{ghisellini2009} remains unchanged while varying L$_{d}$, provided that we scale the solution for $R_{diss}$ with $\sqrt{L_d}$.
Therefore, variations of the disk luminosity in time, and/or systematic errors in the evaluation of disk luminosity from our SDSS+GALEX data (possibly biased by jet emission)
only slightly affect our estimation of $R_{diss}$.\\
The starting point of our modeling is that the emission region is far from the SMBH (at parsec scale), and we motivate this choice with the flat gamma-ray spectrum up to energies of 15 GeV. For 
a different approach and results for other blazars, we refer readers to the work of \cite{tavecchio2010}.  They performed a detailed study
of the localization of the emission region for bright blazars making use of the variability timescale for objects  showing  a spectral cut-off at (10--20)/(1+z) GeV. They obtained that the variability timescale
and spectra are in agreement with a dissipation region inside the BLR.\\
We note that, contrary to the model we used, some authors \citep[e.g., ][ for example]{giommi2011} model the SEDs of both FSRQ and BL Lacs with pure Synchrotron + Synchrotron Self Compton components only.\\ 
Model 1 ($R_{diss}\sim$ 0.2 pc), gives a variability time scale of the order of 3 days, in agreement with the flare duration estimated from the 1-day binned gamma-ray light curve.
This model, however, does not properly reproduce the flat gamma-ray spectrum. In particular, in order to reproduce the observed flux at energies $>$ 10 GeV, it overestimates the spectrum for lower energies. On the contrary,  model 2 ($R_{diss}\sim$7  pc) reproduces the flat-gamma-ray spectrum, but it is not clear whether the predicted variability of the order of $\sim$100 days can be associated with the duration of the gamma-ray activity period estimated from the 1-week/1-month binned gamma-ray light curves.
The third model has been built by relaxing the relation $R_{blob}$=0.1$R_{diss}$ in order to preserve the  variability time-scale estimated from the 1-day gamma-ray light curve (we follow the solution proposed by \citealt{tavecchio2011} for PKS~1222+216), and it still reproduces the gamma-ray spectrum.
Interestingly, the size of the emitting region for PKS 1222+216 ($R_{blob}\sim5\cdot 10^{14}$cm) and for \gbsource\  differ significantly.\\
For the third model, we obtain $R_{blob}$=0.0067$R_{diss}$, in agreement within a factor of two with the prediction of \cite{bromberg2009}, that gives $R_{blob}=10^{-2.5}R_{diss}$, for the case
of efficient conversion of bulk luminosity in radiation in the strong focusing scenario. With the same assumptions, \cite{bromberg2009} assume that the location of the emitting region is at $R_{diss}\sim$2.5 $(L_{jet}/10^{46}$erg s$^{-1}) (R_{BLR}/0.1$ pc$)^{-1}$ pc from the SMBH, where $L_{jet}$ is the jet power.  If we invert this relation, and we make use of our result  ($R_{diss}=4.8$ pc), we obtain $L_{jet}\sim 3.5\cdot 10^{46} erg s^{-1}$.
We must assume that the proton--to--emitting electron ratio is of the order of 0.1 in order to reproduce such a power (in the evaluation of proton power reported in Table \ref{tab:models} we assumed one proton per emitting electron, instead).
We note, however, that \cite{nalewajko2009} evaluated that  efficient radiative conversion could be assumed if the product of the bulk Lorentz factor by the opening angle is $\ga$3, and according to our third model this product is 2.
\section*{Acknowledgements}
The authors wish to acknowledge the anonymous referee for his suggestions.
We are also very grateful to J. Finke for comments and corrections, and for his careful proofreading of several sections of this manuscript.\\
We acknowledge financial contribution from the agreement ASI-INAF I/009/10/0.\\
The {\it AGILE} Mission is funded by the Italian Space Agency (through contract ASI I/089/06/2) with scientific and programmatic participation by
the Italian Institute of Astrophysics (INAF) and the Italian Institute of Nuclear Physics (INFN).\\
This research has made use of data from the MOJAVE database that is maintained by the MOJAVE team \citep{lister2009}.\\
This research has made use of the NASA/IPAC Extragalactic Database (NED) which is operated by the Jet Propulsion Laboratory,
California Institute of Technology, under contract with the National Aeronautics and Space Administration. \\
This research has made use of the SIMBAD database, operated at CDS, Strasbourg, France.\\
RJA is supported by an appointment to the NASA Postdoctoral Program at the Jet Propulsion Laboratory,
administered by Oak Ridge Associated Universities through a contract with NASA.\\
%GG%
The \textit{Fermi} LAT Collaboration acknowledges generous ongoing support
from a number of agencies and institutes that have supported both the
development and the operation of the LAT as well as scientific data analysis.
These include the National Aeronautics and Space Administration and the
Department of Energy in the United States, the Commissariat \`a
l'Energie Atomique
and the Centre National de la Recherche Scientifique / Institut
National de Physique
Nucl\'eaire et de Physique des Particules in France, the Agenzia
Spaziale Italiana
and the Istituto Nazionale di Fisica Nucleare in Italy, the Ministry
of Education,
Culture, Sports, Science and Technology (MEXT), High Energy Accelerator Research
Organization (KEK) and Japan Aerospace Exploration Agency (JAXA) in Japan, and
the K.~A.~Wallenberg Foundation, the Swedish Research Council and the
Swedish National Space Board in Sweden.
Additional support for science analysis during the operations phase is
gratefully
acknowledged from the Istituto Nazionale di Astrofisica in Italy and
the Centre National d'\'Etudes Spatiales in France.\\
%

%\end{harvard}
%
\label{lastpage}

\begin{thebibliography}{10}
%GG%\bibitem[Abdo et al. (2011)]{fermicat2}Abdo A. et al., 2011, ApJ submitted; arXiv:1108.1435;
\bibitem[Abdo et al. (2010)]{abdo3c273}Abdo A. et al., 2010, ApJL, 714, L73-L78;
\bibitem[Adelman-McCharty et al. (2008)]{adelman2008}Adelman McCharty J. K., et al., 2008, ApJS, 175, 297;
\bibitem[Agudo et al. (2011a)]{agudo2011a}Agudo I. et al., 2011, ApJL, 726, 13; 
\bibitem[Agudo et al. (2011b)]{agudo2011b}Agudo I. et al., 2011, ApJL, 735, 10; 
\bibitem[Assef et al. (2011)]{assef2011}Assef R. J., et al., ApJ in press; arXiv:1009.1145;
\bibitem[Atwood et al. (2009)]{atwood2009}Atwood et al., 2009, ApJ, 697, 1071-1102;
\bibitem[Bottcher (2007)]{bottcher2007} Bottcher M., Ap\&SS, 2007, 309, 95-104;
\bibitem[Bromberg \& Levinson (2009)]{bromberg2009} Bromberg O.\& Levinson A., 2009, ApJ, 699, 1274-1280; 
\bibitem[Burrows et al. (2005)]{burrows2005}Burrows D. N., et al. 2005, Space Sci. Rev., 120, 165;
\bibitem[Casandjian \& Grenier (2008)]{casandjian2008}J. M. Casandjian, I. A. Grenier, 2008, A\&A, 489, 849;
\bibitem[Cattaneo et al. (2011)]{cattaneo2011}Cattaneo P. W. et al., 2011, NIMA, 630, 251-257;
\bibitem[Collin et al.(2006)]{collin2006} Collin S. et al., 2006, A\&A, 456, 75;
\bibitem[Dickey \& Lockman (1990)]{dickey1990}Dickey J. M. \& Lockman F. J. , 1990, ARA\&A 28, 215-261;
\bibitem[Denney et al. (2009)]{denney2009}Denney K. D., et al., 2009, ApJ 692, 246;
\bibitem[Denney et al. (2011)]{denney2011}Denney K. D., et al., 2011, PoS (NLSY1), 34;
\bibitem[Esposito et al. (1999)]{esposito1999}Esposito J. A. et al., 1999, ApJS, 123, 203-271;
\bibitem[Finke et al. (2010)]{finke2010}Finke J., et al., 2010, ApJ, 712, 238-249; 
\bibitem[Fitzpatrick (1999)]{Fitzpatrick1999}Fitzpatrick  E. L., 1999, PASP 111, 63-75;
\bibitem[Gehrels et al. (2004)]{Gehrels2004}Gehrels N., et al. 2004, ApJ, 611, 1005;
\bibitem[Ghisellini et al. (2007)]{ghisellini2007} Ghisellini G. et al. 2007, MNRAS, 382, L82-L86:
\bibitem[Ghisellini \& Tavecchio(2009)]{ghisellini2009} Ghisellini G. $\&$ Tavecchio F., 2009, MNRAS, 397, 985-1002;
\bibitem[Ghisellini et al. (2010)]{ghisellini2010}Ghisellini G. et al., 2010, MNRAS, 402, 497;
\bibitem[Ghisellini et al. (2011)]{ghisellinihighz}Ghisellini G. et al., 2011, MNRAS, 411, 911;
\bibitem[Giannios et al. (2010)]{giannios2010}Giannios D. et al., 2010, MNRAS, 402, 1649-1656;
\bibitem[Giommi et al. (2011)]{giommi2011} Giommi P. et al., 2011, A\&A submitted; arXiv:1108.1114v1;
\bibitem[Hartman et al. (1999)]{hartman1999} Hartman R. C., et al., 1999, ApJS, 123, 79-202; 
\bibitem[Ikejiri et al. (2009)]{trispecatel}Ikejiri Y., 2009, Atel 1892;
\bibitem[Ivezic et al. (2007)]{ivezic2007}Ivezic., et al., 2007, ASPC, 364, 165-175; 
\bibitem[Jester et al. (2005)]{jester2005}Jester S., 2005, AJ 130, 873;
\bibitem[Jorstad et al. (2010)]{jorstad2010}Jorstad S. G. et al., 2010, ApJ, 715, 362-384;
\bibitem[Lister et al. (2009)]{lister2009}Lister M. L., et al., 2009, AJ, 137, 3718;
%
\bibitem[Maraschi et al. (1992)]{maraschi1992} Maraschi L., Ghisellini G., and Celotti A., 1992, ApJL, 397, L5-L9;
\bibitem[Marsher \& Bloom (1992)]{marscher1992} Marscher A. P., and Bloom S. D., 1992, Proceedings of The Compton Observatory Science Workshop, 346-353;
\bibitem[Marscher et al. (2010)]{marscher2010}Marscher A. P., et al., 2010, ApJL, 710, 726;
\bibitem[Martin et al. (2005)]{galex2005}Martin D. C. et al., 2005, ApJL, 619, L1-L6; 
\bibitem[Mas-Hesse et al. (2003)]{mas-hesse2003}Mas-Hesse J. M., et al. 2003, A\&A, 411, L261;
\bibitem[Massaro et al. (2010)]{massaro2010}Massaro E., et al., 2010; arXiv:1006.0922;
%
\bibitem[Mattox et al. (1996)]{mattox1996} Mattox J. R., Bertsch, D. L., Chiang, J., et al., 1996, ApJ, 461, 396-407;
%
\bibitem[M\"ucke and Protheroe (2001)]{mucke2001} M\"ucke A., Protheroe R. J., 2001, Astropart. Phys, 15, 121-136;
\bibitem[M\"ucke et al. (2003)]{mucke2003} M\"ucke A. et al, 2003, Astropart. Phys, 18, 593-613;
\bibitem[Nalewajko \& Sikora (2009)]{nalewajko2009}Nalewajko K. \& Sikora M., 2009, MNRAS 392, 1205; 
%
\bibitem[Nolan et al. (2012)]{fermicat2}Nolan P. L. et al., 2012, ApJS, 199, 31;
%
\bibitem[Pacciani et al. (2009)]{pacciani2009}Pacciani L., et al., 2009, A\&A, 494, 49;
\bibitem[Pian et al. (1999)]{pian1999}Pian E. et al., 1999, ApJL, 521, 112;
\bibitem[Pittori et al. (2009)]{pittori2009}Pittori C.,, et al., 2009, A\&A, 506, 1563;
%
\bibitem[Poole et al. (2008)]{Poole08} Poole T. S., Breeveld A. A., Page M. J., et al., 2008, MNRAS, 383, 627
\bibitem[Poutanen \& Stern (2010)]{poutanen2010} Poutanen J \& Stern B., 2010, ApJL, 717, 118;
%
\bibitem[Richards et al.(2002)]{richards2002}Richards G. T., et al., 2002, AJ, 124, 1; 
\bibitem[Roming et al. (2005)]{roming2005}Roming P. W. A., et al. 2005, Space Sci. Rev., 120, 95;
\bibitem[Shakura \& Sunyaev (1973)]{shakura1973}Shakura N. I. \& Sunyaev R. A., 1973, A\&A, 24, 337;
\bibitem[Shaw et al. (2012)]{shaw2012}Shaw M. S., et al., 2012, ApJ, in press;
\bibitem[Sikora et al. (1994)]{sikora1994} Sikora M., Begelman M. C., and Rees M., 1994, ApJ, 421, 153;
\bibitem[Sikora et al. (2008)]{sikora2008} Sikora M., et al., 2008, ApJ, 675, 71-78;
%
\bibitem[Tavani et al. (2009)]{agile} Tavani, M., Barbiellini, G., Argan, A., et al., 2009, A\&A, 502, 995-1013
\bibitem[Tavecchio $\&$ Mazin (2009)]{tavecchio2009} Tavecchio F. $\&$ Mazin D., 2009, MNRAS, 392, L40-L44;
\bibitem[Tavecchio et al. (2010)]{tavecchio2010} Tavecchio F. et al., 2010, MNRAS, 405,L94-L98;
\bibitem[Tavecchio et al. (2011)]{tavecchio2011} Tavecchio F. et al., 2011, A\&A, 534, A86;
\bibitem[Terasranta et al. (2005)]{Terasranta2005}Terasranta H. et Al., 2005, A\&A, 440, 409;
\bibitem[Tramacere et al. (2009)]{tramatel}Tramacere  A., et al., 2009, Atel. 1888;
\bibitem[Ubertini et al. (2003)]{ubertini2003}Ubertini  P., et al., 2003, A\&A, 411, L131-L139;
\bibitem[Urry and Padovani (1995)]{padovani1995} Urry C. M., and Padovani P., 1995, PASP, 107, 803;
\bibitem[Verrecchia et al. (2011)]{verrecchia2011}Verrecchia F. et al., 2011, ASR in press, doi:10.1016/j.asr.2011.05.035;
\bibitem[Vestergaard \& Peterson (2006)]{Vestergaard2006}Vestergaard M., Peterson B. M., 2006, ApJ, 641, 689; 
\bibitem[Watanabe et al. (2005)]{watanabe2005}Watanabe M., et al. 2005, PASP, 117, 870-884;
\bibitem[Winkler et al. (2003)]{winkler2003}Winkler C, et al. 2003, A\&A, 411, L1.
%
\end{thebibliography}
\end{document}